\documentclass[a4]{article}
\usepackage[T1]{fontenc}
\usepackage[utf8]{inputenc}
\usepackage{amsmath, amssymb, amsthm}
\usepackage[english]{babel}
\usepackage{indentfirst}
\usepackage{booktabs}
\usepackage{array}
\usepackage{caption,appendix}
\usepackage{geometry}
\usepackage{float}
\usepackage{cancel}
\usepackage{bbold}
\usepackage{authblk}
\usepackage{empheq}
\begin{document}

\author[1,2,3,4]{Salvatore Capozziello \thanks{capozziello@na.infn.it}}
\author[5,2]{Maurizio Capriolo \thanks{mcapriolo@unisa.it}}
\author[5,6]{Gaetano Lambiase \thanks{glambiase@unisa.it}}





\affil[1]{\emph{Dipartimento di Fisica "E. Pancini", Universit\`a di Napoli {}``Federico II'', Compl. Univ. di Monte S. Angelo, Edificio G, Via Cinthia, I-80126, Napoli, Italy}}
\affil[2]{\emph{Istituto Nazionale di Fisica Nucleare, Sezione  di Napoli, Compl. Univ. di Monte S. Angelo, Edificio G, Via Cinthia, I-80126,  Napoli, Italy}}
\affil[3]{\emph{Scuola Superiore Meridionale, Largo S. Marcellino 10, I-80138, Napoli, Italy}}
\affil[4]{\emph{Department of Mathematics, Faculty of Civil Engineering,VSB-Technical University of Ostrava, Ludvika Podeste 1875/17,  708 00 Ostrava-Poruba, 
\newline
Czech Republic}}
\affil[5]{\emph{Dipartimento di Fisica "E. R. Caianiello",  Universit\`a degli Studi di Salerno, via Giovanni Paolo II, 132 I-84084 Fisciano, Salerno, Italy}}
\affil[6]{\emph{Istituto Nazionale di Fisica Nucleare, Sezione di Napoli, Gruppo Collegato di Salerno, via Giovanni Paolo II, 132 I-84084 Fisciano, Salerno, Italy}}

\title{Energy-Momentum Complex in Higher Order Curvature-Based Local Gravity}
\maketitle

\abstract{An unambiguous definition of gravitational energy remains one of the unresolved issues of physics today. This problem is related to the non-localization of gravitational energy density. In General Relativity, there have been many proposals for defining the gravitational energy density, notably those proposed by Einstein, Tolman, Landau and Lifshitz, Papapetrou, Møller, and Weinberg. In this review, we firstly explored the energy--momentum complex in an $n^{th}$ order gravitational Lagrangian $L=L\left(g_{\mu\nu}, g_{\mu\nu,i_{1}}, g_{\mu\nu,i_{1}i_{2}},g_{\mu\nu,i_{1}i_{2}i_{3}},\cdots, g_{\mu\nu,i_{1}i_{2}i_{3}\cdots i_{n}}\right)$ and then in a gravitational Lagrangian as \mbox{$L_{g}=(\overline{R}+a_{0}R^{2}+\sum_{k=1}^{p} a_{k}R\Box^{k}R)\sqrt{-g}$}. Its gravitational part was obtained by invariance of gravitational action under infinitesimal rigid translations using Noether's theorem. We also showed that this tensor, in general, is not a covariant object but only an affine object, that is, a pseudo-tensor. Therefore, the pseudo-tensor $\tau^{\eta}_{\alpha}$ becomes the one introduced by Einstein if we limit ourselves to General Relativity and its extended corrections have been explicitly indicated. The same method was used to derive the energy--momentum complex in $ f\left (R \right) $ gravity both in Palatini and metric approaches. Moreover, in the weak field approximation the pseudo-tensor $\tau^{\eta}_{\alpha}$ to lowest order in the metric perturbation $h$ was calculated. As a practical application, the power per unit solid angle $\Omega$ emitted by a localized source carried by a gravitational wave in a direction $\hat{x}$ for a fixed wave number $\mathbf{k}$ under a suitable gauge was obtained, through the average value of the pseudo-tensor over a suitable spacetime domain and the local conservation of the pseudo-tensor. As a cosmological application, in a flat Friedmann--Lema\^itre--Robertson--Walker spacetime, the gravitational and matter energy density in $f(R)$ gravity both in Palatini and metric formalism was proposed. The gravitational energy--momentum pseudo-tensor could be a useful tool to investigate further modes of gravitational radiation beyond two standard modes required by General Relativity and to deal with non-local theories of gravity involving $\Box^{-k}$ terms.
}



\section{Introduction}
A widely accepted definition of gravitational energy density and its localization in curved spacetime are serious problems that afflict the general relativity. Several prescriptions for gravitational contribution to energy--momentum density and more generally for energy--momentum complex have been suggested by Einstein, Tolman, Landau and Lifshitz, Papapetrou, Møller, and Weinberg \cite{LL, Einstein, Hasten, GoldbergCL, DBCJ, LLNCLVP, RosenEU, LESSNERM, PalmerGEM, FF}. These attempts are based on the introduction of a super-potential or through the expansion of the Ricci tensor in the metric perturbation $h$. Thus, the gravitational part of the energy--momentum density transforms as an affine tensor not as a covariant tensor, and for this reason, it is not really a tensor but a pseudo-tensor. This affine property of the gravitational stress--energy tensor 
makes the gravitational energy--momentum density not localizable. However, integrating the density over a suitable spatial region at a certain time such as over an asymptotically flat spacetime, viable for isolated systems, the gravitational energy--momentum becomes a four-vector, as meaning that changes in right way under asymptotically flat coordinate transformations. Over all space it becomes quasi independent of the coordinate system, that is, the gravitational energy--momentum of the spacetime exists, but it cannot be localized. In this review a generalization of Einstein's pseudo-tensor to Extended Theories of Gravity \cite{ET,mauro} is proposed by imposing the invariance of the higher order gravitational Lagrangian under an infinitesimal rigid translation and by using Noether's theorem. Then, thanks to a continuity equation, a Noether current and a Noether charge were derived that correspond to a gravitational energy--momentum pseudo-tensor and gravitational energy--momentum, respectively, both locally conserved. 
By weakly perturbing the metric tensor around the Minkowskian metric, a weak-field limit, in a suitable gauge, the gravitational energy--momentum pseudo-tensor for a Lagrangian of $n^{th}$ order appears an object easier to handle. Then, by averaging of the pseudo-tensor over a suitable spacetime domain, it is possible to calculate the power emitted by some localized astrophysical source carried away by the gravitational waves. This approach could be relevant for searching for polarization states of gravitational waves in addition to the two standards of general relativity \cite{greci, patrizia}. Finally, after deriving the gravitational energy--momentum pseudo-tensor in $F(R)$ gravity formulated in Palatini and metric formalism, some cosmological applications were discussed, wherein a flat FLRW metric the total energy density was obtained in both approaches \cite{MWH,vagenas}.

For more details on the issue of energy--momentum localization in modified theories of gravity such as $f(R)$, ${f(R,\Box R,\dots, \Box^{k} R)}$ \cite{vagenas, CCT}, teleparallel gravity and its extended version $f(T)$, see Ref. \cite{Capozziello:2018qcp}. Meanwhile, for a study of wavelike solutions of modified teleparallel gravity necessary for future applications of the pseudo-tensor, see references \cite{CCC1, CCC2}.

The review is organized as follows. Firstly in Sec. \ref{DEFEMTG} some definitions of gravitational pseudo-tensors in general relativity are listed. In Sec. \ref{EMTG} we derived the gravitational energy--momentum pseudo-tensor for a general Lagrangian of $n^{th}$ order through two procedures: the first method uses a variational principle under rigid transformations via Noether's theorem and the second adopts the Landau--Lifshitz procedure \cite{LL} without the introduction of the super-potential. Hence, in Sec. \ref{PNC}, we proved that a stress--energy object is transformed in the correct manner under linear transformations but not under diffeomorphisms and, therefore, is a pseudo-tensor and not a covariant tensor. In Sec. \ref{GEMTR}, we calculated the Euler--Lagrange equations and the gravitational energy--momentum pseudo-tensor for $f(R)$ gravity, always using Noether's theorem applied to a particular one-parameter group of diffeomorphisms given by rigid translations. Therefore, in all models of gravity we obtained the continuity equation for an energy--momentum complex. In Sec. \ref{EMTL}, we derived the gravitational energy--momentum pseudo-tensor of a gravitation field for a particular Lagrangian $L_{g}=(\overline{R}+a_{0}R^{2}+\sum_{k=1}^{p} a_{k}R\Box^{k}R)\sqrt{-g}$. Sec. \ref{EMTLO}, is devoted to the weak-field limit of the gravitational stress--energy pseudo-tensor expanded to lowest order in a small perturbation $h$, i.e., up to $h^2$ order, and we have shown two simple cases where the index $p$ is equal to zero and one. Hence, in Sec. \ref{MVEMT}, we averaged the pseudo-tensor over an suitable region containing the isolated sources and then we found the emitted power carried by the gravitational radiation. Afterward, in Sec. \ref{paltinisection}, in Palatini $f({\cal R})$ gravity, related field equations and related gravitational energy--momentum pseudo-tensor were obtained. Therefore in Sec. \ref{cosmo}, by adopting a flat FLRW spacetime, an explicit calculus of an energy density complex for power law cosmological solutions was performed, also in the metric formalism of $f(R)$. Conclusions are summarized in Sec. \ref{conclusions}. Finally in Appendix~\ref{A}, we proved that the additive terms related to the symmetries of $g_{\mu\nu}$ and its derivatives yield a mean of zero, i.e., $\langle\left(A_{p}\right)_{\alpha}^{\eta}\rangle=\langle\left(B_{p}\right)_{\alpha}^{\eta}\rangle=0$. While in Appendix~\ref{B}, we explicitly showed the six polarization tensors associated with the gravitational waves present in higher-order theories.

\section{Several definitions of gravitational energy--momentum pseudo-tensor in general relativity }\label{DEFEMTG}
Here are some of the most important definitions of gravitational energy--momentum pseudo-tensor in general relativity in the scientific literature, for details see \cite{Xulu}.
\subsection{Einstein energy--momentum complex}
In special relativity the law of conservation of energy and momentum is given by
\begin{equation}\label{conserveinongrav}
\frac{\partial T^{\mu\nu}}{\partial x^{\mu}}=0\ ,
\end{equation}
with $T^{\mu\nu}$ the energy--momentum tensor of matter and non-gravitational fields. In general relativity this principle becomes for general covariance
\begin{equation}\label{conserveigrav}
\nabla_{\mu}T^{\mu\nu}=0\ ,
\end{equation}
which does not correspond to any law of conservation of physical quantities. Einstein therefore formulated the conservation law in the following way
\begin{equation}\label{conserveigravespl}
\frac{\partial \theta_{\mu}^{\phantom{\mu}\nu}}{\partial x^{\nu}}=\frac{\partial}{\partial x^{\nu}}\left(\sqrt{-g}\left(T_{\mu}^{\phantom{\mu}\nu}+t_{\mu}^{\phantom{\mu}\nu}\right)\right)=0\ ,
\end{equation}
where $t_{\mu}^{\phantom{\mu}\nu}$ is an pseudo-tensor. So what is conserved is not only the tensor of non-gravitational fields and matter $T_{\mu}^{\phantom{\mu}\nu}$ but a pseudo-tensor $t_{\mu}^{\phantom{\mu}\nu}$ must be added to it. This pseudo-tensor added can be interpreted as associated with the gravitational field and the energy due to the sum of the contributions of the gravitational fields plus those due to the matter is conserved. However, the pseudo-tensoriality behaviour of $t_{\mu}^{\phantom{\mu}\nu}$ makes it dependent on coordinates and the gravitational energy becomes non localizable. In order to write the Eq.~\eqref{conserveigrav} in the form of an ordinary divergence equation Eq.~\eqref{conserveigravespl}, Einstein starting from the following Lagrangian density which is a non-covariant scalar density
\begin{equation}\label{LagrEinst}
L=\sqrt{-g} g^{\mu\nu}\left(\Gamma^{\sigma}_{\mu\nu}\Gamma^{\rho}_{\sigma\rho}-\Gamma^{\sigma}_{\mu\rho}\Gamma^{\rho}_{\nu\sigma}\right)\ ,
\end{equation}
introduced a pseudo-tensor defined by the relation 
\begin{equation}
\sqrt{-g}t_{\mu}^{\phantom{\mu}\nu}=\frac{1}{16\pi}\left(\delta^{\nu}_{\mu}L-\frac{\partial L}{\partial g^{\rho\sigma}_{\phantom{\rho\sigma},\nu}}g^{\rho\sigma}_{\phantom{\rho\sigma},\mu}\right)\ .
\end{equation}
\subsection{Landau--Lifshitz energy--momentum pseudo-tensor}
The gravitational energy--momentum pseudo-tensor defined by Landau--Lifshitz has the great advantage of being symmetric unlike Einstein's, which in general is not.This allows defining the angular momentum and therefore the related conservation law. We adopt a system of geodetic coordinates where the first derivatives of the metric tensor $g^{\mu\nu}$ vanish. Then, the Eq.~\eqref{conserveigrav} is reduced to \eqref{conserveinongrav} which can be written in terms of the following antisymmetric quantity in the last two indices $\eta^{\mu\nu\sigma}=-\eta^{\mu\sigma\nu}$
\begin{equation}
T^{\mu\nu}=\frac{\partial \eta^{\mu\nu\sigma}}{\partial x^{\sigma}}\ .
\end{equation}
Since the Levi--Civita connection $\Gamma$ vanishes at one point, in such coordinate system it is possible using Einstein’s equations in the presence of matter written in such coordinates, to express the stress--energy tensor of matter $T^{\mu\nu}$ as
\begin{equation}
T^{\mu\nu}=\frac{1}{\left(-g\right)}\frac{\partial}{\partial x^{\sigma}}\left\{\frac{1}{16\pi}\frac{\partial}{\partial x^{\rho}}\left[\left(-g\right)\left(g^{\mu\nu}g^{\sigma\rho}-g^{\mu\sigma}g^{\nu\rho}\right)\right]\right\}\ ,
\end{equation} 
where indicating the term in braces with the antisymmetric quantity in the last two indices
$h^{\mu\nu\sigma}=-h^{\mu\sigma\nu}$, we get
\begin{equation}
\frac{\partial h^{\mu\nu\sigma}}{\partial x^{\sigma}}-\left(-g\right)T^{\mu\nu}=0\ .
\end{equation}
Returning to an arbitrary coordinate system the previous difference does not cancel anymore so we can indicate it with $\left(-g\right)t^{\mu\nu}$ or
\begin{equation}\label{LLECCG}
\left(-g\right)\left(T^{\mu\nu}+t^{\mu\nu}\right)=\frac{\partial h^{\mu\nu\sigma}}{\partial x^{\sigma}}\ .
\end{equation}
Quantities $t^{\mu\nu}$ are symmetric but are not the components of a covariant tensor but affine. Using Einstein’s field equations again it is possible from Eq. \eqref{LLECCG} to get an explicit expression of $t^{\mu\nu}$, defined as the energy--momentum pseudo-tensor of the gravitational field, by means of the derivatives of the components of the metric tensor, that is
\begin{equation}\label{LLEMPT}
\begin{split}
16\pi\left(-g\right)t^{\mu\nu}=&\mathfrak{g}^{\mu\nu}_{\phantom{\mu\nu},\rho}\mathfrak{g}^{\rho\sigma}_{\phantom{\rho\sigma},\sigma}-\mathfrak{g}^{\mu\rho}_{\phantom{\mu\rho},\rho}\mathfrak{g}^{\nu\sigma}_{\phantom{\nu\sigma},\sigma}+\frac{1}{2}g^{\mu\nu}g_{\rho\sigma}\mathfrak{g}^{\rho\alpha}_{\phantom{\rho n},\beta}\mathfrak{g}^{\beta\sigma}_{\phantom{\beta\sigma},\alpha}\\
&-\left(g^{\mu\rho}g_{\sigma\alpha}\mathfrak{g}^{\nu\alpha}_{\phantom{\nu n},\beta}\mathfrak{g}^{\sigma\beta}_{\phantom{\sigma\beta},\rho}+g^{\nu\rho}g_{\sigma \alpha}\mathfrak{g}^{\mu\alpha}_{\phantom{\mu\alpha},\beta}\mathfrak{g}^{\sigma\beta}_{\phantom{\sigma\beta},\rho}\right)+g_{\rho\sigma}g^{\alpha\beta}\mathfrak{g}^{\mu\rho}_{\phantom{\mu\rho},\alpha}\mathfrak{g}^{\nu\sigma}_{\phantom{\nu m},\beta}\\
&+\frac{1}{8}\left(2g^{\mu\rho}g^{\nu\sigma}-g^{\mu\nu}g^{\rho\sigma}\right)\left(2g_{\alpha\beta}g_{\gamma\lambda}-g_{\beta \gamma}g_{\alpha\lambda}\right)\mathfrak{g}^{\alpha\lambda}_{\phantom{\alpha\lambda},\rho}\mathfrak{g}^{\beta\gamma}_{\phantom{\beta\gamma},\sigma}\ ,
\end{split}
\end{equation}
where $\mathfrak{g}^{\mu\nu}=\sqrt{-g}g^{\mu\nu}$. 
\subsection{M\o ller energy--momentum complex}
The energy--momentum pseudo-tensors $t^{\mu\nu}$ of both Einstein and Landau--Lifshitz besides having the flaw of being tensors only affine and not covariant also depend on the choice of coordinates. Then, M\o ller looked for an expression for energy and gravitational momentum independent of the particular coordinate system. To do this M\o ller exploited the fact that the pseudo-tensor including matter plus gravity $\theta^{\mu\nu}=T^{\mu\nu}+t^{\mu\nu}$ can be defined at less than a magnitude $S^{\mu\nu}$ at zero divergence $\partial_{\mu} S^{\mu\nu}=0$. In 1958 M\o ller proposed the following complex tensor of energy--momentum complex $\mathcal{T}_{\mu}^{\phantom{\mu}\nu}=\theta_{\mu}^{\phantom{\mu}\nu}+S_{\mu}^{\phantom{\mu}\nu}$ looking for the $S_{\mu}^{\phantom{\mu}\nu}$ such that $\mathcal{T}_{\mu}^{\phantom{\mu}\nu}$ transformed as a tensor for only spatial transformations
\begin{equation}\label{EMTGM}
\mathcal{T}_{\mu}^{\phantom{\mu}\nu}=\frac{1}{8\pi}\partial_{\rho}\left[\sqrt{-g}\left(g_{\mu\sigma,\lambda}-g_{\mu\lambda,\sigma}\right)g^{\lambda\nu}g^{\sigma\rho}\right]\ ,
\end{equation}
where the expression in square brackets is the antisymmetric super-potential $U_{\mu}^{\phantom{\mu}\nu\rho}=-U_{\mu}^{\phantom{\mu}\rho\nu}$ such that
\begin{equation}
\partial_{\nu}\mathcal{T}_{\mu}^{\phantom{\mu}\nu}=0\ .
\end{equation}

\subsection{Papapetrou energy--momentum pseudo-tensor}
Papapetrou in 1948 used the generalized Belifante method to derive his pseudo-tensor due to the complex of energy--momentum contributions using Tolman’s expression of Einstein's total pseudo-tensor $\theta_{\mu}^{\phantom{\mu}\nu}$ (\ref{conserveigravespl}) i.e.
\begin{equation}
\theta_{\mu}^{\phantom{\mu}\nu}=\frac{1}{8\pi}\frac{\partial}{\partial x^{\rho}}\left(-\mathfrak{g}^{\nu\sigma}\frac{\partial L}{\partial \mathfrak{g}^{\mu\sigma}_{\phantom{\mu\sigma},\rho}}+\frac{1}{2}\delta_{\mu}^{\nu}\mathfrak{g}^{\alpha\beta}\frac{\partial L}{\partial \mathfrak{g}^{\alpha\beta}_{\phantom{\alpha\beta},\rho}}\right)\ ,
\end{equation}
where $L$ is Einstein Lagrangian give by \eqref{LagrEinst} and $\mathfrak{g}^{\nu\sigma}$ have been defined in Eq.~\eqref{LLEMPT}. Belifante’s method consists in finding a symmetric quantity $\Omega^{\mu\nu}=\Omega^{\nu\mu}$ divergence free which differs by $\eta^{\mu\rho}\theta_{\rho}^{\phantom{\rho}\nu}$ only for an antisymmetric quantity divergence in the first two indices $B^{\mu\nu\rho}=-B^{\mu\nu\rho}$ i.e.
\begin{equation}
\Omega^{\mu\nu}=\eta^{\mu\rho}\theta_{\rho}^{\phantom{\rho}\nu}+\frac{\partial}{\partial x^{\rho}}B^{\mu\nu\rho}\ ,
\end{equation}
such that
\begin{equation}\label{coserlocpapa}
\frac{\partial}{\partial x^{\nu}}\Omega^{\mu\nu}=0\ ,
\end{equation}
with $\eta_{\mu\nu}=diag\left(1,-1,-1,-1\right)$. Expressing $B^{\mu\nu\rho}$ in terms of of the field spin density $S^{\mu\nu\rho}$
\begin{equation}
B^{\mu\nu\rho}=-\frac{1}{2}\left(S^{\mu\nu\rho}+S^{\rho\mu\nu}+S^{\rho\nu\mu}\right)\ ,
\end{equation}
you get after a few counts the expression for the total pseudo-tensor $\Omega^{\mu\nu}$
\begin{equation}
\Omega^{\mu\nu}=\frac{1}{16\pi}\frac{\partial^{2}}{\partial x^{\rho}x^{\sigma}}\left[\sqrt{-g}\left(g^{\mu\nu}\eta^{\rho\sigma}-g^{\mu\rho}\eta^{\nu\sigma}-g^{\rho\sigma}\eta{\mu\nu}-g^{\nu\sigma}\eta^{\mu\rho}\right)\right]\ .
\end{equation}
This geometric object is symmetric with respect to the first two indices$\mu$ e $\nu$.

\subsection{ Weinberg gravitational energy--momentum pseudo-tensor}
Weinberg \cite{WE} derived the gravitational energy--momentum pseudo-tensor by adopting a quasi-minkowskian coordinate system. In this system the metric tensor $g_{\mu\nu}$ tends to that of Minkowski $\eta_{\mu\nu}$ at great distances from a localized material system. We write the metric $g_{\mu\nu}$ as the sum of the metric of Minkowski $\eta_{\mu\nu}$ plus $h_{\mu\nu}$ that goes to zero to infinity 
\begin{equation}
g_{\mu\nu}=\eta_{\mu\nu}+h_{\mu\nu}\ .
\end{equation}
We linearize Einstein equations $G_{\mu\nu}=-8\pi G T_{\mu\nu}$, expanding Ricci tensor $R_{\mu\nu}$ in terms of powers of $h_{\mu\nu}$ as
\begin{equation}\label{EELFIRST}
R^{\left(1\right)}_{\phantom{\left(1\right)}\mu\nu}-\frac{1}{2}\eta_{\mu\nu}R^{\left(1\right)}=-8\pi G\left[T_{\mu\nu}+t_{\mu\nu}\right]\ ,
\end{equation}
where 
\begin{equation}
t_{\mu\nu}=\frac{1}{8\pi G}\left(R_{\mu\nu}-\frac{1}{2}g_{\mu\nu}R-R^{\left(1\right)}_{\phantom{\left(1\right)}\mu\nu}+\frac{1}{2}\eta_{\mu\nu}R^{\left(1\right)}\right)\ ,
\end{equation}
is the gravitational energy--momentum pseudo-tensor. So in the Eq.~\eqref{EELFIRST} you see that reading the equation from right to left, $t_{\mu\nu}$ assumes the meaning of the source of the linearized curvature together with the tensor of the non-gravitational fields and of the matter $T_{\mu\nu}$. From the linearized Bianchi law to which quantity $R^{\left(1\right)}_{\phantom{\left(1\right)}\mu\nu}$ must satisfy, we get the following local conservation law
\begin{equation}
\frac{\partial}{\partial x^{\nu}}\left(T^{\mu\nu}+t^{\mu\nu}\right)=0\ .
\end{equation}
The pseudo-tensor $t_{\mu\nu}$ to the second order in $h$ is
\begin{equation}
t_{\mu\nu}=\frac{1}{8\pi G}\left(-\frac{1}{2}h_{\mu\nu}R^{\left(1\right)}+\frac{1}{2}\eta_{\mu\nu}\eta^{\rho\sigma}R^{\left(1\right)}_{\phantom{\left(1\right)}\rho\sigma}+R^{\left(2\right)}_{\phantom{\left(2\right)}\mu\nu}-\frac{1}{2}\eta_{\mu\nu}\eta^{\rho\sigma}R^{\left(2\right)}_{\phantom{\left(2\right)}\rho\sigma}\right)+\mathcal{O}\left(h^{3}\right)\ ,
\end{equation}
where $R_{\mu\nu}$ to first order in $h$ is
\begin{equation}
R^{\left(1\right)}_{\phantom{\left(1\right)}\mu\nu}=\frac{1}{2}\left(\frac{\partial^{2}h^{\lambda}_{\phantom{\lambda}\lambda}}{\partial x^{\mu}\partial x^{\nu}}-\frac{\partial^{2}h^{\lambda}_{\phantom{\lambda}\mu}}{\partial x^{\lambda}\partial x^{\nu}}-\frac{\partial^{2}h^{\lambda}_{\phantom{\lambda}\nu}}{\partial x^{\lambda}\partial x^{\mu}}+\frac{\partial^{2}h_{\mu\nu}}{\partial x^{\lambda}\partial x_{\lambda}}\right)\ ,
\end{equation}
while to second order $h$ becomes
\begin{equation}
\begin{split}
R^{\left(2\right)}_{\phantom{\left(2\right)}\mu\nu}=&-\frac{1}{2}h^{\lambda\rho}\left(\frac{\partial^{2}h_{\lambda\rho}}{\partial x^{\nu}\partial x^{\mu}}-\frac{\partial^{2}h_{\mu\rho}}{\partial x^{\nu}\partial x^{\lambda}}-\frac{\partial^{2}h_{\lambda\nu}}{\partial x^{\rho}\partial x^{\mu}}+\frac{\partial^{2}h_{\mu\nu}}{\partial x^{\rho}\partial x^{\lambda}}\right)\\
&+\frac{1}{4}\left(2\frac{\partial h^{\rho}_{\phantom{\rho}\sigma}}{\partial x^{\rho}}-\frac{\partial h^{\rho}_{\phantom{\rho}\rho}}{\partial x^{\sigma}}\right)\left(\frac{\partial h^{\sigma}_{\phantom{\sigma}\mu}}{\partial x^{\nu}}+\frac{\partial h^{\sigma}_{\phantom{\sigma}\nu}}{\partial x^{\mu}}-\frac{\partial h_{\mu\nu}}{\partial x_{\sigma}}\right)\\
&-\frac{1}{4}\left(\frac{\partial h_{\sigma\nu}}{\partial x^{\lambda}}+\frac{\partial h_{\sigma\lambda}}{\partial x^{\nu}}-\frac{\partial h_{\lambda\nu}}{\partial x^{\sigma}}\right)\left(\frac{\partial h^{\sigma}_{\phantom{\sigma}\mu}}{\partial x_{\lambda}}+\frac{\partial h^{\sigma\lambda}}{\partial x^{\mu}}-\frac{\partial h^{\lambda}_{\phantom{\lambda}\mu}}{\partial x_{\sigma}}\right)
\end{split}\ .
\end{equation}

\section{Energy--Momentum Complex in curvature based gravity}

 \subsection{The gravitational energy--momentum ''tensor'' of $n^{th}$ order Lagrangian}\label{EMTG}
Let us examine the energy--momentum complex for a fourth order gravitational Lagrangian, that is, which depends up to fourth derivatives of the metric tensor $g_{\mu\nu}$ as $L=L(g_{\mu\nu}, g_{\mu\nu,\rho}, g_{\mu\nu,\rho\lambda},g_{\mu\nu,\rho\lambda\xi}, g_{\mu\nu,\rho\lambda\xi\sigma})$, whose field equations, in general, are of eighth order in metric formalism (see also \cite{arturo1,stelle}). In this manner we include all possible curvature invariants, not only $\Box$ operators, into the gravitational action. Then, we will generalize the approach to a gravitational Lagrangian of $n$-th order, i.e., which depends up to $n^{th}$ derivatives of metric tensor. We will derive the energy--momentum tensor using the Noether's theorem, imposing that gravitational action is invariant under global translations \cite{LL}. In this review the metric signature of $g_{\mu\nu}$ adopted is $(+\ \ , -\ \ , -\ \ , -)$, while Ricci tensor is defined as
$R_{\mu\nu}=R_{\ \ \mu\rho\nu}^{\rho}$ and Riemann tensor as $R_{\ \ \beta\mu\nu}^{\alpha}=\Gamma_{\beta\nu,\mu}^{\alpha}+\ldots$.

Let us vary the gravitational action with respect to metric $g_{\mu\nu}$ and coordinates $x^{\mu}$ \cite{ET, PML, FQ}
\begin{equation}\label{1}
I=\int_{\Omega}d^{4}x L \rightarrow \tilde{\delta} I=\int_{\Omega^{\prime}} d^{4}x^{\prime} L^{\prime}-\int_{\Omega}d^{4}x L=\int_{\Omega}d^{4}x \left[{\delta}L+\partial_{\mu}\left(L\delta x^{\mu}\right)\right]\ ,
\end{equation}
where $\tilde{\delta}$ stands for the local variation while $\delta$ means the total variation, that is, keeping the value of coordinate $x$ fixed. By infinitesimal transformations as
\begin{equation}\label{2}
x^{\prime\mu}=x^{\mu}+\epsilon^{\mu}\left(x\right)\ ,
\end{equation}
the total variation of the metric tensor reads
\begin{equation}\label{3}
\delta g_{\mu\nu}=g^{\prime}_{\mu\nu}\left(x\right)-g_{\mu\nu}\left(x\right)=-\epsilon^{\alpha}\partial_{\alpha}g_{\mu\nu}-g_{\mu\alpha}\partial_{\nu}\epsilon^{\alpha}-g_{\nu\alpha}\partial_{\mu}\epsilon^{\alpha}\ .
\end{equation}
 Under global transformation, $\partial_{\lambda}\epsilon^{\mu}=0$, the functional variation of the metric becomes $\delta g_{\mu\nu}=-\epsilon^{\alpha}\partial_{\alpha}g_{\mu\nu}$. If we also require that the action to be invariant under this transformation, that is, $\tilde{\delta I} =0$, from arbitrariness of domain of integration $\Omega$, we have
\begin{equation}\label{4}
\begin{split}
 0=\delta L +\partial_{\mu}\left(L\delta x^{\mu}\right)=\biggl(\frac{\partial L}{\partial g_{\mu\nu}}-\partial_{\rho}\frac{\partial L}{\partial g_{\mu\nu,\rho}}+\partial_{\rho}\partial_{\lambda}\frac{\partial L}{\partial g_{\mu\nu,\rho\lambda}}-\partial_{\rho}\partial_{\lambda}\partial_{\xi}
 \frac{\partial L}{\partial g_{\mu\nu,\rho\lambda\xi}}\\+\partial_{\rho}\partial_{\lambda}\partial_{\xi}\partial_{\sigma}\frac{\partial L}{\partial g_{\mu\nu,\rho\lambda\xi\sigma}}\biggr)\delta g_{\mu\nu}+\partial_{\eta}\left(2\chi\sqrt{-g}\tau_{\alpha}^{\eta}\right)\epsilon^{\alpha}\ ,
\end{split}
\end{equation}
where the explicit expression of gravitational energy--momentum tensor, that we will see being a pseudo-tensor or affine tensor, is
\begin{multline}\label{7}
\tau_{\alpha}^{\eta}=\frac{1}{2\chi\sqrt{-g}}\biggl[\left(\frac{\partial L}{\partial g_{\mu\nu,\eta}}-\partial_{\lambda}\frac{\partial L}{\partial g_{\mu\nu,\eta\lambda}}+\partial_{\lambda}\partial_{\xi}\frac{\partial L}{\partial g_{\mu\nu,\eta\lambda\xi}}-\partial_{\lambda}\partial_{\xi}\partial_{\sigma}\frac{\partial L}{\partial g_{\mu\nu,\eta\lambda\xi\sigma}}\right)g_{\mu\nu,\alpha}\\
+\left(\frac{\partial L}{\partial g_{\mu\nu,\rho\eta}}-\partial_{\xi}\frac{\partial L}{\partial g_{\mu\nu,\rho\eta\xi}}+\partial_{\xi}\partial_{\sigma}\frac{\partial L}{\partial g_{\mu\nu,\rho\eta\xi\sigma}}\right)g_{\mu\nu,\alpha\rho}
+\left(\frac{\partial L}{\partial g_{\mu\nu,\rho\lambda\eta}}-\partial_{\sigma}\frac{\partial L}{\partial g_{\mu\nu,\rho\lambda\eta\sigma}}\right)g_{\mu\nu,\rho\lambda\alpha}\\
+\frac{\partial L}{\partial g_{\mu\nu,\rho\lambda\eta\sigma}}g_{\mu\nu,\rho\lambda\xi\alpha}-\delta^{\eta}_{\alpha}L\biggr]\ .
\end{multline}
If the metric tensor $g_{\mu\nu}$ satisfies the Euler--Lagrange equations for our gravitational Lagrangian
\begin{equation}\label{5}
\frac{\delta L}{\delta g_{\mu\nu}}=\frac{\partial L}{\partial g_{\mu\nu}}-\partial_{\rho}\frac{\partial L}{\partial g_{\mu\nu,\rho}}+\partial_{\rho}\partial_{\lambda}\frac{\partial L}{\partial g_{\mu\nu,\rho\lambda}}-\partial_{\rho}\partial_{\lambda}\partial_{\xi}\frac{\partial L}{\partial g_{\mu\nu,\rho\lambda\xi}}+\partial_{\rho}\partial_{\lambda}\partial_{\xi}\partial_{\sigma}\frac{\partial L}{\partial g_{\mu\nu,\rho\lambda\xi\sigma}}=0\ ,
\end{equation}
for an arbitrary $\epsilon^{\alpha}$, we get a local continuity equation for our Noether current 
\begin{equation}\label{6}
\partial_{\eta}\left(\sqrt{-g} \tau_{\alpha}^{\eta}\right)=0\ .
\end{equation}
In a more compact form, the gravitational energy--momentum tensor takes the following form
\begin{multline}\label{tensem4}
\tau_{\alpha}^{\eta}=\frac{1}{2\chi\sqrt{-g}}\biggl[\sum_{m=0}^{3}\left(-1\right)^{m}\left(
\frac{\partial L}{\partial g_{\mu\nu,\eta i_{0}\cdots i_{m}}}\right)_{,i_{0}\cdots i_{m}}g_{\mu\nu,\alpha}
 \\ +\sum_{j=0}^{2}\sum_{m=j+1}^{3}\left(-1\right)^{j}\left(
\frac{\partial L}{\partial g_{\mu\nu,\eta i_{0}\cdots i_{m}}}\right)_{,i_{0}\cdots i_{j}}g_{\mu\nu,i_{j+1}\cdots i_{m}\alpha}-\delta_{\alpha}^{\eta}L\biggr]\ ,
\end{multline}
where we used the following notation
\begin{equation*}
\left(\right)_{,i_{0}}=\mathbb{I} ; \qquad \left(\right)_{,i_{0}\cdots i_{m}}=
\begin{cases}
\left(\right)_{,i_{1}}& \quad \text{if} \quad m=1\\
\left(\right)_{,i_{1}i_{2}}& \quad \text{if} \quad m=2\\
\left(\right)_{,i_{1}i_{2}i_{3}}& \quad \text{if} \quad m=3\\
\text{and so on}&
\end{cases}
;\qquad \left(\right)_{,i_{k}\.i_{k}}=\left(\right)_{,i_{k}}
\end{equation*}
Let us now generalize our approach considering a general Lagrangian density depending up to $n^{th}$ derivative of $g_{\mu\nu}$, that is, $L=L\left(g_{\mu\nu}, g_{\mu\nu,i_{1}}, g_{\mu\nu,i_{1}i_{2}},g_{\mu\nu,i_{1}i_{2}i_{3}},\cdots, g_{\mu\nu,i_{1}i_{2}i_{3}\cdots i_{n}}\right)$. Total variation of Lagrangian $L$ and its Euler--Lagrange equations yield
\begin{equation}\label{8}
 \delta L=\sum_{m=0}^{n}\frac{\partial L}{\partial g_{\mu\nu,i_{0}\cdots i_{m}}}\delta g_{\mu\nu,i_{0}\cdots i_{m}}=\sum_{m=0}^{n}\frac{\partial L}{\partial g_{\mu\nu,i_{0}\cdots i_{m}}}\partial_{i_{0}\cdots i_{m}}\delta g_{\mu\nu}\ ,
\end{equation}
\begin{equation}\label{9}
 \frac{\delta L}{\delta g_{\mu\nu}}=\sum_{m=0}^{n}\left(-1\right)^{m}\partial_{i_{0}\cdots i_{m}}\frac{\partial L}{\partial g_{\mu\nu,i_{0}\cdots i_{m}}}=0\ ,
\end{equation}
where $\delta/\delta g_{\mu\nu}$ is the functional derivative, while it is possible to exchange the variation $\delta$ with the derivatives $\delta g_{\mu\nu,i_{0}\cdots i_{m}}=\partial_{i_{0}\cdots i_{m}}\delta g_{\mu\nu}$, because we are varying keeping $x$ fixed. So, we can find a most general local continuity equation which allows us to define the energy--momentum pseudo-tensor (which is an affine tensor as it will be proved later) for the gravitational field of $2n^{th}$ order gravity 
\begin{multline}\label{tensemN}
\tau_{\alpha}^{\eta}=\frac{1}{2\chi\sqrt{-g}}\biggl[\sum_{m=0}^{n-1}\left(-1\right)^{m}\left(
\frac{\partial L}{\partial g_{\mu\nu,\eta i_{0}\cdots i_{m}}}\right)_{,i_{0}\cdots i_{m}}g_{\mu\nu,\alpha}\\+\Theta_{\left[2,+\infty\right[}\left(n\right)\sum_{j=0}^{n-2}\sum_{m=j+1}^{n-1}\left(-1\right)^{j}\left(
\frac{\partial L}{\partial g_{\mu\nu,\eta i_{0}\cdots i_{m}}}\right)_{,i_{0}\cdots i_{j}}g_{\mu\nu,i_{j+1}\cdots i_{m}\alpha}-\delta_{\alpha}^{\eta}L\biggr]\ ,
\end{multline}
where $\Theta$ is the Heaviside function
\begin{equation}
\Theta_{\left[a,+\infty\right[}\left(n\right)=
\begin{cases}
1& \quad \text{if} \quad n\in \left[a,+\infty\right[\\
0& \quad \text{otherwise}
\end{cases}\ .
\end{equation}
If fields and its derivatives vanish on boundary of our spatial region or rapidly decreasing to the spatial infinite on an infinity spacelike hypersurface, the gravitational energy--momentum tensor is totally conserved and satisfies a more general conservation law. 
An alternative way to obtain the tensor \eqref{tensemN} is the procedure developed by Landau \cite{LL}. For example, we start by deriving the tensor \eqref{tensem4}, because its generalization to higher order Lagrangians is the same. First of all, let us impose the stationary condition and vary the action with respect to the metric to find the field equations under the hypothesis that both $\delta g_{\mu\nu}$ and the variation of derivative $\delta \partial^{n}g$ vanish on the boundary of integration domain, canceling the surface integrals. Hence, the following occurs:
\begin{gather}\label{15}
\delta I=\delta\int_{\Omega}d^{4}x L\left(g_{\mu\nu}, g_{\mu\nu,\rho}, g_{\mu\nu,\rho\lambda},g_{\mu\nu,\rho\lambda\xi}, g_{\mu\nu,\rho\lambda\xi\sigma}\right) =0\ , \\
\updownarrow\nonumber\\
\frac{\partial L}{\partial g_{\mu\nu}}-\partial_{\rho}\frac{\partial L}{\partial g_{\mu\nu,\rho}}+\partial_{\rho}\partial_{\lambda}\frac{\partial L}{\partial g_{\mu\nu,\rho\lambda}}-\partial_{\rho}\partial_{\lambda}\partial_{\xi}\frac{\partial L}{\partial g_{\mu\nu,\rho\lambda\xi}}+\partial_{\rho}\partial_{\lambda}\partial_{\xi}\partial_{\sigma}\frac{\partial L}{\partial g_{\mu\nu,\rho\lambda\xi\sigma}}=0\ .
\end{gather}
Now, we perform the derivative of Lagrangian respect to metric tensor and then we put it into the field equations \eqref{15}. We obtain
\begin{multline}\label{16}
\frac{\partial L}{\partial x^{\alpha}} =\frac{\partial L}{\partial g_{\mu\nu}}\frac{\partial g_{\mu\nu}}{\partial x^{\alpha}}+\frac{\partial L}{\partial g_{\mu\nu,\rho}}\frac{\partial g_{\mu\nu,\rho}}{\partial x^{\alpha}}+\frac{\partial L}{\partial g_{\mu\nu,\rho\lambda}}\frac{\partial g_{\mu\nu,\rho\lambda}}{\partial x^{\alpha}}\\
+\frac{\partial L}{\partial g_{\mu\nu,\rho\lambda\xi}}\frac{\partial g_{\mu\nu,\rho\lambda\xi}}{\partial x^{\alpha}}+\frac{\partial L}{\partial g_{\mu\nu,\rho\lambda\xi\sigma}}\frac{\partial g_{\mu\nu,\rho\lambda\xi\sigma}}{\partial x^{\alpha}}\\
 =\partial_{\rho} \frac{\partial L}{\partial g_{\mu\nu,\rho}}g_{\mu\nu,\alpha}-\partial_{\rho}\partial_{\lambda}\frac{\partial L}{\partial g_{\mu\nu,\rho\lambda}}g_{\mu\nu,\alpha}+\partial_{\rho}\partial_{\lambda}\partial_{\xi}\frac{\partial L}{\partial g_{\mu\nu,\rho\lambda\xi}}g_{\mu\nu,\alpha}-\partial_{\rho}\partial_{\lambda}\partial_{\xi}\partial_{\sigma}\frac{\partial L}{\partial g_{\mu\nu,\rho\lambda\xi\sigma}}g_{\mu\nu,\alpha}\\
+\frac{\partial L}{\partial g_{\mu\nu,\rho}} g_{\mu\nu,\rho\alpha}+\frac{\partial L}{\partial g_{\mu\nu,\rho\lambda}}g_{\mu\nu,\rho\lambda\alpha}+\frac{\partial L}{\partial g_{\mu\nu,\rho\lambda\xi}}g_{\mu\nu,\rho\lambda\xi\alpha}+\frac{\partial L}{\partial g_{\mu\nu,\rho\lambda\xi\sigma}} g_{\mu\nu,\rho\lambda\xi\sigma\alpha}\\
=\partial_{\rho}\left(\frac{\partial L}{\partial g_{\mu\nu,\rho}}g_{\mu\nu,\alpha}\right)-\partial_{\rho}\left(\partial_{\lambda}\frac{\partial L}{\partial g_{\mu\nu,\rho\lambda}}g_{\mu\nu,\alpha}\right)+\partial_{\lambda}\left(\frac{\partial L}{\partial g_{\mu\nu,\rho\lambda}}g_{\mu\nu,\rho\alpha}\right)\\
+\partial_{\rho}\left(\partial_{\lambda}\partial_{\xi}\frac{\partial L}{\partial g_{\mu\nu,\rho\lambda\xi}}g_{\mu\nu,\alpha}\right)
+\partial_{\lambda}\left(\frac{\partial L}{\partial g_{\mu\nu,\rho\lambda\xi}}g_{\mu\nu,\rho\xi\alpha}\right)\\
-\partial_{\xi}\left(\partial_{\lambda}\frac{\partial L}{\partial g_{\mu\nu,\rho\lambda\xi}}g_{\mu\nu,\alpha\rho}\right)-\partial_{\rho}\left(\partial_{\lambda}\partial_{\xi}\partial_{\sigma}\frac{\partial L}{\partial g_{\mu\nu,\rho\lambda\xi\sigma}}g_{\mu\nu,\alpha}\right)\\
+\partial_{\lambda}\left(\frac{\partial L}{\partial g_{\mu\nu,\rho\lambda\xi\sigma}}g_{\mu\nu,\rho\xi\sigma\alpha}\right)-\partial_{\xi}\left(\partial_{\lambda}\frac{\partial L}{\partial g_{\mu\nu,\rho\lambda\xi\sigma}}g_{\mu\nu,\rho\sigma\alpha}\right)
\\+\partial_{\sigma}\left(\partial_{\xi}\partial_{\lambda}\frac{\partial L}{\partial g_{\mu\nu,\rho\lambda\xi\sigma}}g_{\mu\nu,\rho\alpha}\right)\ .
\end{multline}
Grouping together terms and renaming dumb indices, we obtain
\begin{equation}\label{17}
\partial_{\eta}\left(\sqrt{-g}\tau^{\eta}_{\alpha}\right)=0\ ,
\end{equation}
that is, the pseudo-tensor is locally conserved, where $\tau^{\eta}_{\alpha}$ is the tensor defined in \eqref{tensem4}. 

The energy--momentum complex, instead, can be derived considering the material Lagrangian $L_{m}=2\chi\sqrt{-g}\mathcal{L}_{m}$ with stress--energy tensor given by
\begin{equation}\label{18}
T^{\eta\alpha}=\frac{2}{\sqrt{-g}}\frac{\delta \left(\sqrt{-g}\mathcal{L}_{m}\right)}{\delta g_{\eta\alpha}}\ .
\end{equation}
Thus, we use the field equations in presence of matter, namely 
\begin{equation}\label{19}
P^{\eta\alpha}=\chi T^{\eta\alpha}\ ,
\end{equation} 
where
\begin{equation}\label{20}
P^{\eta\alpha}=-\frac{1}{\sqrt{-g}}\frac{\delta L_{g}}{\delta g_{\eta\alpha}}\,\qquad \mbox{with the coupling} \quad \chi=\frac{8\pi G}{c^{4}}\ .
\end{equation}
By field equations~\eqref{19}, we obtain
\begin{multline}\label{21}
\left(2\chi\sqrt{-g}\tau^{\eta}_{\alpha}\right)_{,\eta}=-\sqrt{-g}P^{\rho\sigma}g_{\rho\sigma,\alpha}=-\chi\sqrt{-g}T^{\rho\sigma}g_{\rho\sigma,\alpha}\\
=2\chi\sqrt{-g} T^{\eta}_{\alpha;\eta}-\left(2\chi\sqrt{-g} T^{\eta}_{\alpha}\right)_{,\eta}\ ,
\end{multline}
\begin{equation}\label{22}
\partial_{\eta}\left[\sqrt{-g}\left(\tau^{\eta}_{\alpha}+ T^{\eta}_{\alpha}\right)\right]=\sqrt{-g}T^{\eta}_{\alpha;\eta}\ ,
\end{equation}
being
\begin{multline}\label{23}
\delta L +\partial_{\mu}\left(L\delta x^{\mu}\right)=-P^{\mu\nu}\sqrt{-g}\delta g_{\mu\nu}+\partial_{\eta}\left(2\chi\sqrt{-g}\tau^{\eta}_{\alpha}\right)\epsilon^{\alpha}\\
=\left[\sqrt{-g}P^{\mu\nu}g_{\mu\nu,\alpha}+\partial_{\eta}\left(2\chi\sqrt{-g}\tau^{\eta}_{\alpha}\right)\right]\epsilon^{\alpha}=0\ ,
\end{multline}
and because from symmetry of tensor $T^{\eta}_{\alpha}$, one gets 
\begin{equation}\label{24}
\sqrt{-g}T^{\eta}_{\alpha;\eta}=\left(\sqrt{-g}T^{\eta}_{\alpha}\right)_{,\eta}-\frac{1}{2}g_{\rho\sigma,\alpha}T^{\rho\sigma}\sqrt{-g}\ .
\end{equation}
The relation \eqref{22} tells us that the conservation law of the energy--momentum complex, i.e., the sum of two stress--energy tensors due to matter plus gravitational fields, is related to the covariant derivative of the only matter part. From contracted Bianchi identities we get the total conservation law and conversely
\begin{equation}\label{eqcontgeneralizzata}
G^{\eta\alpha}_{;\eta}=0 \leftrightarrow P^{\eta\alpha}_{;\eta}=0 \leftrightarrow T^{\eta\alpha}_{;\eta}=0 \leftrightarrow \partial_{\eta}\left[\sqrt{-g}\left(\tau^{\eta}_{\alpha}+ T^{\eta}_{\alpha}\right)\right]=0\ ,
\end{equation}
where ${\displaystyle G^{\eta\alpha}=R^{\eta\alpha}-\frac{1}{2}g^{\eta\alpha}R}$ is the Einstein tensor and the locally conserved energy--momentum complex is given by 
\begin{equation}
\mathcal{T}_{\alpha}^{\eta}=\sqrt{-g}\left(\tau^{\eta}_{\alpha}+ T^{\eta}_{\alpha}\right)\ .
\end{equation}
In a nutshell, the contracted Bianchi identities lead to the local conservation of energy--momentum complex or, viceversa, the local conservation of matter and gravitational fields involves the contracted Bianchi identities (see also \cite{CF} for a detailed discussion in modified gravity).

From the local continuity equation~\eqref{eqcontgeneralizzata}, it is possible to derive some conserved quantities, Noether charges, such as the total 4-momentum of matter plus gravitational field. If we require that the metric tensor derivatives up to the $n^{th}$ order vanish on the 3-dimensional space-domain $\Sigma$, the surface integrals are zero over the boundary $\partial\Sigma$, that is 
\begin{equation}\label{25}
\partial_{0}\int_{\Sigma}d^{3}x \sqrt{-g}\left(T^{\mu0}+\tau^{\mu0}\right)=-\int_{\partial\Sigma}d\sigma_{i} \sqrt{-g}\left(T^{\mu i}+\tau^{\mu i}\right)=0\ ,
\end{equation} 
where $\Sigma$ is a slice of 4-dimensional manifold of spacetime at $t$ fixed and $\partial\Sigma$ its boundary. Such conditions are fulfilled by when we are in the presence of localized objects, where we can take a spatial domain that becomes flat to infinity, i.e, a asymptotically flat spacetime. So, the energy and total momentum conserved become Ref. \cite{MTW}
\begin{equation}\label{26}
 P^{\mu}=\int_{\Sigma}d^{3}x \sqrt{-g}\left(T^{\mu0}+\tau^{\mu0}\right)\ .
\end{equation}
These quantities are very useful in astrophysical applications \cite{Straumann}.

\subsection{Non-covariance of gravitational energy--momentum tensor}
\label{PNC}
We will prove that the tensor $\tau^{\eta}_{\alpha}$ is not a covariant object but affine, that is, it is changes like a tensor under affine transformations \cite{WP}, i.e., a pseudo-tensor. We will limit ourselves first to a particular case, $n=2$, where the tensor \eqref{tensemN} reads 
\begin{equation}\label{27}
\tau^{\eta}_{\alpha}=\frac{1}{2\chi\sqrt{-g}}\left[\left(\frac{\partial L}{\partial g_{\mu\nu,\eta}}-\partial_{\lambda}\frac{\partial L}{\partial g_{\mu\nu,\eta\lambda}}\right)g_{\mu\nu,\alpha}+\frac{\partial L}{\partial g_{\mu\nu,\eta\xi}}g_{\mu\nu,\xi\alpha}-\delta^{\eta}_{\alpha} L\right]\ ,
\end{equation}
We will show that, while under a general diffeomorphism transformation $x^{\prime}=x^{\prime}\left(x\right)$, the tensor changes as
\begin{equation}\label{28}
\tau^{\prime\eta}_{\ \alpha}\left(x^{\prime}\right) \neq \text{J}^{\eta}_{\sigma}\text{J}^{-1\tau}_{\ \ \ \alpha}\tau^{\sigma}_{\tau}\left(x\right)\ ,
\end{equation}
with Jacobian matrix and determinant defined as 
\begin{equation}\label{29}
\text{J}^{\eta}_{\sigma}=\frac{\partial x^{\prime\eta}}{\partial x^{\sigma}}\qquad \text{J}^{-1\tau}_{\ \ \ \alpha}=\frac{\partial x^{\tau}}{\partial x^{\prime\alpha}}\qquad \text{det}\left(\text{J}^{\alpha}_{\beta}\right)=\vert J \vert =\frac{1}{\text{J}^{-1}}\ ,
\end{equation}
under the following affine transformations 
\begin{equation}\label{30}
x^{\prime\mu}=\Lambda^{\mu}_{\nu}x^{\nu}\qquad \text{J}^{\mu}_{\nu}=\Lambda^{\mu}_{\nu} \qquad \vert \Lambda \vert \neq 0\ ,
\end{equation}
the tensor is transformed as
\begin{equation}\label{31}
\tau^{\prime\eta}_{\ \alpha}\left(x^{\prime}\right)=\Lambda^{\eta}_{\sigma}\Lambda^{-1\tau}_{\ \ \ \alpha}\tau^{\sigma}_{\tau}\left(x\right)\ .
\end{equation}
Generally, following identities occur
\begin{equation*}
\begin{split}
\sqrt{-g^{\prime}}&=\sqrt{-g}\qquad\qquad \ \ \qquad\quad \, \,\text{where $g$ is a scalar density of weight $w=-2$ }\ ,\\
L^{\prime}&=\text{J}^{-1}L\qquad\qquad \ \qquad \qquad \text{where $L$ is a scalar density of weight $w=-1$ }\ ,\\
g^{\prime}_{\mu\nu,\alpha}\left(x^{\prime}\right)&=\text{J}^{-1 a}_{\ \ \ \mu}\text{J}^{-1 b}_{\ \ \ \nu}\text{J}^{-1 c}_{\ \ \ \alpha}g_{ab,c}\left(x\right)+\partial^{\prime}_{\alpha}\left[\text{J}^{-1 a}_{\ \ \ \mu}\text{J}^{-1 b}_{\ \ \ \nu}\right]g_{ab}\left(x\right)\ ,\\
\frac{\partial g_{\gamma\rho,\tau}}{\partial g^{\prime}_{\mu\nu,\eta}}&=\frac{1}{2}\left[\left(\delta_{a}^{\mu}\delta_{b}^{\nu}+\delta_{a}^{\nu}\delta_{b}^{\mu}\right)\delta_{c}^{\eta}\right]\text{J}_{\gamma}^{a}\text{J}_{\rho}^{b}\text{J}_{\tau}^{c}=\text{J}^{(\mu}_{\gamma}\text{J}^{\nu)}_{\rho}\text{J}^{\eta}_{\tau}\ ,\\
\frac{\partial L^{\prime}}{\partial g^{\prime}_{\mu\nu,\eta}}&=\text{J}^{-1}\text{J}^{(\mu}_{\gamma}\text{J}^{\nu)}_{\rho}\text{J}^{\eta}_{\tau}\frac{\partial L}{\partial g_{\gamma\rho,\tau}}=\text{J}^{-1}\text{J}^{\mu}_{\gamma}\text{J}^{\nu}_{\rho}\text{J}^{\eta}_{\tau}\frac{\partial L}{\partial g_{\gamma\rho,\tau}}\ \\
& \qquad\qquad\qquad\qquad\qquad\qquad \text{tensorial density (3,0) of weight $w=-1$}\ ,\\
g^{\prime}_{\mu\nu,\xi\alpha}\left(x^{\prime}\right)&=\text{J}^{-1 a}_{\ \ \ \mu}\text{J}^{-1 b}_{\ \ \ \nu}\text{J}^{-1 c}_{\ \ \ \alpha}\text{J}^{-1 d}_{\ \ \ \xi}g_{ab,cd}\left(x\right)+\partial^{\prime 2}_{\xi\alpha}\left[\text{J}^{-1 a}_{\ \ \ \mu}\text{J}^{-1 b}_{\ \ \ \nu}\right]g_{ab}\left(x\right)\\
&+\partial^{\prime}_{\alpha}\left[\text{J}^{-1 a}_{\ \ \ \mu}\text{J}^{-1 b}_{\ \ \ \nu}\right]\text{J}^{-1 d}_{\ \ \ \xi}g_{ab,d}\left(x\right)+\partial^{\prime}_{\xi}\left[\text{J}^{-1 a}_{\ \ \ \mu}\text{J}^{-1 b}_{\ \ \ \nu}\text{J}^{-1 c}_{\ \ \ \alpha}\right]g_{ab,c}\left(x\right)\ ,\\
\frac{\partial g_{\gamma\rho,\tau\epsilon}}{\partial g_{\mu\nu,\eta\xi}^{\prime}}&=\left(\delta_{a}^{(\mu}\delta_{b}^{\nu)}\delta_{c}^{(\eta}\delta_{d}^{\xi)}\right)\text{J}_{\gamma}^{a}\text{J}_{\rho}^{b}\text{J}_{\tau}^{c}\text{J}_{\epsilon}^{d}=\text{J}_{\gamma}^{(\mu}\text{J}_{\rho}^{\nu)}\text{J}_{\tau}^{(\eta}\text{J}_{\epsilon}^{\xi)}\ ,\\
\frac{\partial L^{\prime}}{\partial g^{\prime}_{\mu\nu,\eta\xi}}&=\text{J}^{-1}\text{J}^{(\mu}_{\gamma}\text{J}^{\nu)}_{\rho}\text{J}^{(\eta}_{\tau}\text{J}^{\xi)}_{\epsilon}\frac{\partial L}{\partial g_{\gamma\rho,\tau\epsilon}}=\text{J}^{-1}\text{J}^{\mu}_{\gamma}\text{J}^{\nu}_{\rho}\text{J}^{\eta}_{\tau}\text{J}^{\xi}_{\epsilon}\frac{\partial L}{\partial g_{\gamma\rho,\tau\epsilon}}\\
 & \qquad\qquad\qquad\qquad\qquad\qquad\text{tensorial density (4,0) of weight $w=-1$}\ ,\\
\partial^{\prime}_{\lambda}\frac{\partial L^{\prime}}{\partial g^{\prime}_{\mu\nu,\eta\lambda}}&=\text{J}^{-1}\text{J}^{\mu}_{\gamma}\text{J}^{\nu}_{\rho}\text{J}^{\eta}_{\tau}\text{J}^{\lambda}_{\epsilon}\text{J}^{-1 \sigma}_{\ \ \ \lambda}\partial_{\sigma}\frac{\partial L}{\partial g_{\gamma\rho,\tau\epsilon}}+\partial^{\prime}_{\lambda}\left[\text{J}^{-1}\text{J}^{\mu}_{\gamma}\text{J}^{\nu}_{\rho}\text{J}^{\eta}_{\tau}\text{J}^{\lambda}_{\epsilon}\right]\frac{\partial L}{\partial g_{\gamma\rho,\tau\epsilon}}\ ,
\end{split}
\end{equation*}
and by symmetry of $B_{\alpha\beta}$, i.e., $B_{\alpha\beta}=B_{\beta\alpha}$ follows that $A^{(\alpha\beta)}B_{\alpha\beta}=A^{\alpha\beta}B_{\alpha\beta}$.
Then we have
\begin{equation*}
\frac{\partial L^{\prime}}{\partial g^{\prime}_{\mu\nu,\eta}}g^{\prime}_{\mu\nu,\alpha}=\text{J}^{-1}\text{J}^{\eta}_{\tau}\text{J}^{-1\pi}_{\ \ \ \alpha}\frac{\partial L}{\partial g_{\gamma\rho,\tau}}g_{\gamma\rho,\pi}\left(x\right)+\frac{\partial}{\partial x^{\prime\alpha}}\left[\text{J}^{-1a}_{\ \ \ \mu}\text{J}^{-1b}_{\ \ \ \nu}\right]g_{ab}\left(x\right)\text{J}^{-1}\text{J}^{\mu}_{\gamma}\text{J}^{\nu}_{\rho}\text{J}^{\eta}_{\tau}\frac{\partial L}{\partial g_{\gamma\rho,\tau}}\ ,\\
\end{equation*}
\begin{multline*}
\partial^{\prime}_{\lambda}\frac{\partial L^{\prime}}{\partial g^{\prime}_{\mu\nu,\eta\lambda}}g^{\prime}_{\mu\nu,\alpha}\left(x^{\prime}\right)=\text{J}^{-1}\text{J}^{\eta}_{\tau}\text{J}^{-1c}_{\ \ \ \alpha}\partial_{\sigma}\frac{\partial L}{\partial g_{ab,\tau\sigma}}g_{ab,c}+\partial^{\prime}_{\lambda}\left[\text{J}^{-1}\text{J}^{\mu}_{\gamma}\text{J}^{\nu}_{\rho}\text{J}^{\eta}_{\tau}\text{J}^{\lambda}_{\epsilon}\right]
\partial^{\prime}_{\alpha}\left[\text{J}^{-1a}_{\ \ \ \mu}\text{J}^{-1b}_{\ \ \ \nu}\right]g_{ab}\left(x\right)\frac{\partial L}{\partial g_{\gamma\rho,\tau\epsilon}}\\
+\text{J}^{-1}\text{J}^{\mu}_{\gamma}\text{J}^{\nu}_{\rho}\text{J}^{\eta}_{\tau}\partial_{\sigma}\frac{\partial L}{\partial g_{\gamma\rho,\tau\sigma}}\partial^{\prime}_{\alpha}\left[\text{J}^{-1a}_{\ \ \ \mu}\text{J}^{-1b}_{\ \ \ \nu}\right]g_{ab}+\partial^{\prime}_{\lambda}\left[\text{J}^{-1}\text{J}^{\mu}_{\gamma}\text{J}^{\nu}_{\rho}\text{J}^{\eta}_{\tau}\text{J}^{\lambda}_{\epsilon}\right]\text{J}^{-1a}_{\ \ \ \mu}\text{J}^{-1b}_{\ \ \ \nu}\text{J}^{-1c}_{\ \ \ \alpha}\frac{\partial L}{\partial g_{\gamma\rho,\tau\epsilon}}g_{ab,c} \ ,
\end{multline*}
\begin{multline*}
\frac{\partial L^{\prime}}{\partial g^{\prime}_{\mu\nu,\eta\xi}}g^{\prime}_{\mu\nu,\xi\alpha}\left(x^{\prime}\right)=\text{J}^{-1}\text{J}^{\eta}_{\tau}\text{J}^{-1\omega}_{\ \ \ \alpha}\frac{\partial L}{\partial g_{\gamma\rho,\tau\epsilon}}g_{\gamma\rho,\omega\epsilon}\left(x\right)+\text{J}^{-1}\partial^{\prime 2}_{\xi\alpha}\left[\text{J}^{-1a}_{\ \ \ \mu}\text{J}^{-1b}_{\ \ \ \nu}\right]g_{ab}\left(x\right)\text{J}^{\mu}_{\gamma}\text{J}^{\nu}_{\rho}\text{J}^{\eta}_{\tau}\text{J}^{\xi}_{\epsilon}\frac{\partial L}{\partial g_{\gamma\rho,\tau\epsilon}}\\
+\text{J}^{-1}\partial^{\prime}_{\alpha}\left[\text{J}^{-1a}_{\ \ \ \mu}\text{J}^{-1b}_{\ \ \ \nu}\right]g_{ab,d}\left(x\right)\text{J}^{\mu}_{\gamma}\text{J}^{\nu}_{\rho}\text{J}^{\eta}_{\tau}\frac{\partial L}{\partial g_{\gamma\rho,\tau d}}+\text{J}^{-1}\text{J}^{\mu}_{\gamma}\text{J}^{\nu}_{\rho}\text{J}^{\eta}_{\tau}\text{J}^{\xi}_{\epsilon}\partial^{\prime}_{\xi}\left[\text{J}^{-1a}_{\ \ \ \mu}\text{J}^{-1b}_{\ \ \ \nu}\text{J}^{-1c}_{\ \ \ \alpha}\right]g_{ab,c}\left(x\right)\frac{\partial L}{\partial g_{\gamma\rho,\tau\epsilon}},
\end{multline*}
Finally, taking into account previous relations we get
\begin{equation}\label{affinitatensore}
 \tau^{\prime\eta}_{\ \alpha}\left(x^{\prime}\right)=\text{J}^{\eta}_{\sigma}\text{J}^{-1\tau}_{\ \ \ \alpha}\tau^{\sigma}_{\tau}\left(x\right)+\left\{\text{terms containing }\frac{\partial^{2} x}{\partial x^{\prime 2}},\frac{\partial^{3} x}{\partial x^{\prime 3}} \right\}\ .
\end{equation}
Extra terms that include derivatives of order equal to or greater than two vanish for each non-singular affine transformation but not for generic diffeomorphisms. This proves that gravitational stress--energy tensor is non-covariant but affine, that is, it is invariant under affine transformations due to non-covariance of the derivatives of the metric tensor $g_{\mu\nu}$, that make it at least affine.
Generalizing to $n$-th order Lagrangian, metric tensor derivatives change as 
\begin{multline}\label{32}
 g^{\prime}_{\mu\nu,i_{1}\cdots i_{m}\alpha}\left(x^{\prime}\right)=\text{J}^{-1\alpha}_{\ \ \ \mu}\text{J}^{-1\beta}_{\ \ \ \nu}\text{J}^{-1 j_{1}}_{\ \ \ i_{1}}\cdots\text{J}^{-1 j_{m}}_{\ \ \ i_{m}}\text{J}^{-1\tau}_{\ \ \ \alpha}g_{\alpha\beta,j_{1}\cdots j_{m}\tau}\left(x\right)\\
+\left\{\text{containing terms}\;\frac{\partial^{2} x}{\partial x^{\prime 2}},\cdots,\frac{\partial^{m+2}x}{\partial x^{\prime m+2}} \right\},
\end{multline}
and derivatives of Lagrangian as 
\begin{equation*}
 \frac{\partial L^{\prime}}{\partial g^{\prime}_{\mu\nu,\eta i_{0}\cdots i_{m}}}=\text{J}^{-1}\text{J}^{\mu}_{\gamma}\text{J}^{\nu}_{\rho}\text{J}^{\eta}_{\tau}\text{J}^{i_{1}}_{j_{1}}\cdots \text{J}^{i_{m}}_{j_{m}} \frac{\partial L}{\partial g_{\gamma\rho,\tau j_{1}\cdots j_{m}}}\quad\text{tensorial density (m+3,0) of weight $w=-1$}\ ,
\end{equation*}
so that the non covariance of tensor $\tau^{\eta}_{\alpha}$ appears. Otherwise, we obtain for affine transformations
\begin{equation*}
\frac{\partial^{2}x}{\partial x^{\prime 2}}=\cdots=\frac{\partial^{m+2}x}{\partial x^{\prime m+2}}=0\ ,
\end{equation*}
\begin{equation*}
 \tau^{\prime\eta}_{\ \alpha}\left(x^{\prime}\right)=\Lambda^{\eta}_{\sigma}\Lambda^{-1\tau}_{\ \ \ \alpha}\tau^{\sigma}_{\tau}\left(x\right)\ ,
\end{equation*}
that is, the energy--momentum tensor of gravitational field is a pseudo-tensor. This result generalize to Extended Theories of Gravity the result in \cite{LL}. The affine character of the stress--energy tensor $\tau ^ {\eta} _ {\alpha}$ is a exhibition of the non localizability of gravitational energy density. Specifically, the gravitational energy in a finite-dimensional space, at a given time, depends on the choice of coordinate system \cite{DI,MTW}. 
It is worth highlighting that the existence of particular Lagrangians for which extra terms in Eq.~\eqref{32} vanish cannot be excluded a priori. This is because terms depending on derivatives in the bracket \eqref{affinitatensore} such as ${\frac{\partial^{2} x}{\partial x^{\prime 2}},\cdots,\frac{\partial^{m+2}x}{\partial x^{\prime m+2}}}$, can cancel each other out. Consequently, the energy--momentum pseudo-tensor ${\tau_{\alpha}^{\eta}}$ become a covariant tensor. However, due to the structure of \eqref{affinitatensore}, in general, ${\tau_{\alpha}^{\eta}}$ is a pseudo-tensor.
\subsection{The gravitational energy--momentum pseudo-tensor of $f\left(R\right)$ gravity}\label{GEMTR}
Let us examine the gravitational stress--energy tensor in the $f\left(R\right)$ gravity. Now, the gravitational action is given by
\begin{equation}\label{33}
\mathcal{S}_{f\left(R\right)}=\frac{1}{2\kappa^{2}}\int_{\Omega}d^{4}x\sqrt{-g}f\left(R\right)\,
\end{equation}
with the coupling $\kappa^{2}=8\pi G/c^{4}$. We perform the variation $\tilde{\delta}$ with respect to the metric $g^{\mu\nu}$ and coordinates $x^{\mu}$ for a generic infinitesimal transformation
\begin{equation}\label{34}
x^{\prime\mu}=x^{\mu}+\delta x^{\mu}\,\qquad g^{\prime\mu\nu}\left(x^{\prime}\right)= g^{\mu\nu}\left(x\right)+\tilde{\delta}g^{\mu\nu}\,\qquad g^{\prime\mu\nu}\left(x\right)= g^{\mu\nu}\left(x\right)+\delta g^{\mu\nu}\,
\end{equation}
\begin{equation}\label{varloc}
\tilde{\delta}\mathcal{S}_{f\left(R\right)}=\frac{1}{2\kappa^{2}}\int_{\Omega}d^{4}x\left[\delta\left(\sqrt{-g}f\left(R\right)\right)+\partial_{\mu}\left(\sqrt{-g}f\left(R\right) \delta x^{\mu}\right)\right]\,
\end{equation}
where $\delta$ is the global variation keeping $x$ fixed.
Thus, we get \cite{ET, CF, HE, SGMM, OR, BR}
\begin{multline}\label{varlocfR}
\tilde{\delta}\mathcal{S}_{f\left(R\right)}=\frac{1}{2\kappa^{2}}\int_{\Omega}d^{4}x\sqrt{-g}\Biggl[f'\left(R\right)R_{\mu\nu}-\frac{1}{2}g_{\mu\nu}f\left(R\right)-\nabla_{\mu}\nabla_{\nu}f'\left(R\right)\\
+g_{\mu\nu}\Box f'\left(R\right)\Biggr]\delta g^{\mu\nu}+\int_{\Omega}d^{4}x\partial_{\alpha}\Biggl\{\frac{\sqrt{-g}}{2\kappa^{2}}\biggl[\partial_{\beta}f'\left(R\right)\left(g^{\eta\rho}g^{\alpha\beta}-g^{\alpha\eta}g^{\rho\beta}\right)\delta g_{\eta\rho}\\
+f'\left(R\right)\Bigl[\left(\stackrel{\circ}\Gamma{}^{\rho\eta\alpha}-\stackrel{\circ}\Gamma{}^{\eta\sigma}{}_{\sigma}g^{\alpha\rho}\right)\delta g_{\eta\rho}+\left(g^{\alpha\eta}g^{\tau\rho}-g^{\eta\rho}g^{\alpha\tau}\right)\delta g_{\eta\rho,\tau}\Bigr]+f\left(R\right)\delta_{\lambda}^{\alpha}\delta x^{\lambda}\biggr]\Biggr\}\,
\end{multline}
where $f'\left(R\right)=\partial f/\partial R$. By the condition of stationarity of the action at $x$ fixed, that is, $\delta\mathcal{S}_{f\left(R\right)}=0$, in a given domain $\Omega$ where the total variation of both metric and its first derivatives are zero on the boundary, that is, $\delta g_{\mu\nu}\vert_{\partial\Omega}=0$ and $\delta \left(\partial_{\alpha}g_{\mu\nu}\right)\vert_{\partial\Omega}=0$, the field equations in vacuum become 
\begin{multline}\label{ECFRV}
P^{f\left(R\right)}_{\mu\nu}=\frac{2\kappa^{2}}{\sqrt{-g}}\frac{\delta L_{f\left(R\right)}}{\delta g^{\mu\nu}}\\
=f'\left(R\right)R_{\mu\nu}-\frac{1}{2}g_{\mu\nu}f\left(R\right)-\nabla_{\mu}\nabla_{\nu}f'\left(R\right)+g_{\mu\nu}\Box f'\left(R\right)=0\,
\end{multline}
where $2\kappa^{2}L_{f\left(R\right)}=\sqrt{-g}f\left(R\right)$. For an infinitesimal transformation such as a rigid translation, one gets
\begin{equation}\label{TrRig}
x^{\prime\mu}=x^{\mu}+\epsilon^{\mu}\Rightarrow \delta g_{\mu\nu}=-\epsilon^{\lambda}g_{\mu\nu,\lambda}\, 
\end{equation}
because $\partial_{\mu} \epsilon^{\mu}=0$. When the local variation of the action vanishes and the field $g_{\mu\nu}$ fulfils the field equations, we obtain the continuity equation

\begin{equation}\label{35}
\tilde{\delta}\mathcal{S}_{f\left(R\right)}=0 \Rightarrow \partial_{\sigma}\left(\sqrt{-g}\tau^{\sigma}_{\phantom{\sigma}{\lambda | f\left(R\right)}}\right)=0\,
\end{equation}
where \emph{the gravitational energy--momentum pseudo-tensor of $f\left(R\right)$ gravity} is defined as 

\begin{multline}\label{36}
2\kappa^{2}\tau^{\sigma}{}_{\lambda | f\left(R\right)}=2\partial_{\beta}f'\left(R\right)g^{\eta[\rho}g^{\sigma]\beta}g_{\eta\rho,\lambda}\\+f'\left(R\right)\Bigl[\bigl(\stackrel{\circ}{\Gamma}{}^{\rho\eta\sigma}
-\stackrel{\circ}{\Gamma}{}^{\eta\alpha}_{\phantom{\eta\alpha}{\alpha}}g^{\sigma\rho}\bigr)g_{\eta\rho,\lambda}+2g^{\sigma[\eta}g^{\tau]\rho}g_{\eta\rho,\tau\lambda}\Bigr]-f\left(R\right)\delta_{\lambda}^{\sigma}\ ,
\end{multline}
with $\stackrel{\circ}{\Gamma}{}^{\rho\eta\sigma}=g^{\eta\epsilon}g^{\sigma\varphi}\stackrel{\circ}{\Gamma}{}^{\rho}_{\phantom{\eta}{\epsilon\varphi}}$, and $\stackrel{\circ}{\Gamma}{}^{\eta\alpha}_{\phantom{\eta\alpha}{\alpha}}=g^{\alpha\epsilon}\stackrel{\circ}{\Gamma}{}^{\eta}_{\phantom{\eta}{\epsilon\alpha}}$.
Now to derive an equation of continuity for energy--momentum complex, we must also include matter fields, as in matter action
\begin{equation}\label{37}
\mathcal{S}_{m}=\int_{\Omega}d^{4}x L_{m}\,
\end{equation}
where $L_{m}$ depends, at most, on first derivatives of metric $g_{\mu\nu}$. Varying the matter action \eqref{37}, it gets
\begin{equation}\label{38}
\delta\mathcal{S}_{m}=\int_{\Omega}d^{4}x \frac{\delta L_{m}}{\delta g^{\mu\nu}}\delta g^{\mu\nu}=\int_{\Omega}d^{4}x \left(\frac{\sqrt{-g}}{2}\right)T^{\left(m\right)}_{\mu\nu}\delta g^{\mu\nu}\,
\end{equation}
where the energy--momentum tensor of matter fields $T^{\left(m\right)}_{\mu\nu}$ is defined as
\begin{equation}\label{39}
T^{\left(m\right)}_{\mu\nu}=-\frac{2}{\sqrt{-g}}\frac{\delta L_{m}}{\delta g^{\mu\nu}}\,.
\end{equation}
So minimizing the total action $\mathcal{S_{T}}=\mathcal{S}_{f\left(R\right)}+\mathcal{S}_{m}$ and imposing suitable boundary conditions, field equations in presence of matter take the following form
\begin{equation}\label{ECFRM}
{}^{f\left(R\right)}P_{\mu\nu}=\kappa^{2}T^{\left(m\right)}_{\mu\nu}\,.
\end{equation}
According to contracted Bianchi identities and following formula 
\begin{equation}\label{trinabla}
\nabla^{\nu}\nabla_{\mu}\nabla_{\nu} f\left(R\right)=R^{\alpha}_{\phantom{\alpha}\mu}\nabla_{\alpha}f\left(R\right)+\nabla_{\mu}\Box f\left(R\right)
\end{equation}
we derive equivalences 
\begin{equation}\label{IBFR}
\nabla^{\nu}G_{\mu\nu}=0\leftrightarrow\nabla^{\nu} \left({}^{f\left(R\right)}P_{\mu\nu}\right)=0\leftrightarrow\nabla^{\nu}T^{\left(m\right)}_{\mu\nu}=0\,
\end{equation}\label{40}
The variation \eqref{varlocfR} of gravitational action, the rigid translation \eqref{TrRig} and the matter field equations \eqref{ECFRM} give
\begin{align}\label{41}
\delta L_{f\left(R\right)}+\partial_{\sigma}\left(L_{f\left(R\right)}\delta x^{\sigma}\right)&=\frac{\sqrt{-g}}{2\kappa^{2}}P_{{f\left(R\right)}}^{\mu\nu}\delta g_{\mu\nu}-\partial_{\sigma}\left(\sqrt{-g}\tau^{\sigma}_{\phantom{\sigma}{\lambda}}\right)\epsilon^{\lambda}\nonumber\\&=\left[-\frac{1}{2}\sqrt{-g}T_{\left(m\right)}^{\mu\nu}g_{\mu\nu,\lambda}-\partial_{\sigma}\left(\sqrt{-g}\tau^{\sigma}_{\phantom{\sigma}{\lambda}}\right)\right]\epsilon^{\lambda}\,.
\end{align}
Taking into account Eq.~\eqref{24}, the expression Eq.~\eqref{41} yields
\begin{equation}\label{42}
\delta L_{f\left(R\right)}+\partial_{\sigma}\left(L_{f\left(R\right)}\delta x^{\sigma}\right)=\left[-\partial_{\sigma}\left(\sqrt{-g}T^{\sigma}_{\phantom{\sigma}{\lambda}}\right)+\sqrt{-g}T^{\sigma}_{\phantom{\sigma}{\lambda;\sigma}}-\partial_{\sigma}\left(\sqrt{-g}\tau^{\sigma}_{\phantom{\sigma}{\lambda}}\right)\right]\epsilon^{\lambda}\,.
\end{equation}
Imposing the local variation to zero, under rigid translations, we have 
\begin{equation}\label{43}
\delta L_{f\left(R\right)}+\partial_{\sigma}\left(L_{f\left(R\right)}\delta x^{\sigma}\right)=0 \rightarrow \partial_{\sigma}\left[\sqrt{-g}\left(\tau^{\sigma}_{\ \lambda}+T^{\sigma}{}_{\lambda}\right)\right]=\sqrt{-g}\nabla_{\sigma}T^{\sigma}{}_{\lambda}\,.
\end{equation}
From the contracted Bianchi identities \eqref{IBFR}, we derive local conservation law for the energy--momentum complex $\mathcal{T}^{\sigma}_{\phantom{\sigma}\lambda}$ in $f(R)$ gravity
\begin{equation}\label{44}
\partial_{\sigma}\left[\sqrt{-g}\left(\tau^{\sigma}_{\phantom{\sigma}{\lambda | f\left(R\right)}}+T^{\sigma}_{\phantom{\sigma}{\lambda}}\right)\right]=0\,.
\end{equation} 
where
\begin{equation}\label{45}
\mathcal{T}^{\sigma}_{\phantom{\sigma}\lambda}=\sqrt{-g}\left(\tau^{\sigma}_{\phantom{\sigma}{\lambda | f\left(R\right)}}+T^{\sigma}_{\phantom{\sigma}{\lambda}}\right)\ .
\end{equation}

\subsection{The gravitational energy--momentum pseudo-tensor of higher order gravity}\label{EMTL}
Let us now address theories of gravity of order higher than fourth, where terms containing $\Box$ operators occur in the action up to $n$ times. In supergravity and, more broadly, in gauge theories concerning with gravity \cite{Modesto, Modesto1, Briscese}, these theories are not only effective field theories, but also fundamental theories. Actually, there is at least a subclass of local higher derivative theories, the so called Lee--Wick theories, that are unitary and super-renormalizable or finite at quantum level as demonstrated in \cite{modesto1, modesto2}. Then, we consider the linear and quadratic part of the Ricci scalar $R$, the first $\overline{R}$ depends only on first derivative of metric tensor $g_{\mu\nu}$ and the second $R^{\star}$ depends linearly on second derivative of metric tensor, as follows \cite{LL,WP,WE}
\begin{equation}\label{46}
 R=R^{\star}+\overline{R}\ ,
\end{equation}
\begin{equation}\label{47}
 R^{\star}=g^{\mu\nu}\left(\Gamma^{\rho}_{\mu\nu,\rho}-\Gamma^{\rho}_{\mu\rho,\nu}\right)\ ,
\end{equation}
\begin{equation}\label{48}
\overline{R}=g^{\mu\nu}\left(\Gamma_{\mu\nu}^{\sigma}\Gamma_{\sigma\rho}^{\rho}-\Gamma_{\mu\sigma}^{\rho}\Gamma_{\nu\rho}^{\sigma}\right)\ .
\end{equation}
Hence, we want to derive the energy--momentum pseudo-tensor $\tau_{\alpha}^{\eta}$ for a gravitational Lagrangian given by
\begin{equation}
\label{higher}
L_{g}=(\overline{R}+a_{0}R^{2}+\sum_{k=1}^{p} a_{k}R\Box^{k}R)\sqrt{-g}\ ,
\end{equation}
that has been first considered in \cite{Quandt}. 
Therefore, for the purpose of derive the pseudo-tensor $\tau^{\eta}_{\alpha}$, we have first to calculate derivatives present into the Eq. \eqref{tensemN}, namely
\begin{align}
 \frac{\partial L}{\partial g_{\mu\nu,\eta}}&=\sqrt{-g}\left[\frac{\partial\overline{R}}{\partial g_{\mu\nu,\eta}}+\left(2a_{0}R+\sum_{k=1}^{p}a_{k}\Box^{k}R\right)\frac{\partial R}{\partial g_{\mu\nu,\eta}}+\sum_{k=1}^{p}a_{k}R\frac{\partial\Box^{k}R}{\partial g_{\mu\nu,\eta}}\right]\ ,\\
 -\partial_{\lambda}\left(\frac{\partial L}{\partial g_{\mu\nu,\eta\lambda}}\right)&=-\partial_{\lambda}\left(\sqrt{-g}\left[\left(2a_{0}R+\sum_{k=1}^{p}a_{k}\Box^{k}R\right)\frac{\partial R}{\partial g_{\mu\nu,\eta\lambda}}+\sum_{k=1}^{p}a_{k}R\frac{\partial\Box^{k}R}{\partial g_{\mu\nu,\eta\lambda}}\right]\right)\ ,
\end{align}
\begin{multline}\bar{49}
\sum_{m=2}^{n-1}\left(-1\right)^{m}\left(\frac{\partial L}{\partial g_{\mu\nu,\eta i_{0}\cdots i_{m}}}\right)_{,i_{0}\cdots i_{m}}=\sum_{m=2}^{n-1}\sum_{k=1}^{p}\left(-1\right)^{m}\partial_{i_{0}\cdots i_{m}}\left[\sqrt{-g}a_{k}R\frac{\partial\Box^{k}R}{\partial g_{\mu\nu,\eta i_{0}\cdots i_{m}}}\right]\\
=\sum_{k=1}^{p}\sum_{m=2}^{2p+3}\left(-1\right)^{m}\partial_{i_{0}\cdots i_{m}}\left[\sqrt{-g}a_{k}R\frac{\partial\Box^{k}R}{\partial g_{\mu\nu,\eta i_{0}\cdots i_{m}}}\right]\\
=\sum_{k=1}^{p}\sum_{m=2}^{2k+1}\left(-1\right)^{m}\partial_{i_{0}\cdots i_{m}}\left[\sqrt{-g}a_{k}R\frac{\partial\Box^{k}R}{\partial g_{\mu\nu,\eta i_{0}\cdots i_{m}}}\right]\ ,
\end{multline}
where $\lambda=i_{1}$, $n=2p+4$ and
\begin{equation}\label{50}
\frac{\partial\Box^{k}R}{\partial g_{\mu\nu,\eta i_{0}\cdots i_{m}}}=0 \qquad \text{if}\quad m>2k+1\ .
\end{equation}
Then, after algebraic manipulations, one have
\begin{multline}\label{51}
\sum_{j=0}^{n-2}\sum_{m=j+1}^{n-1}\left(-1\right)^{j}\left(\frac{\partial L}{\partial g_{\mu\nu,\eta i_{0}\cdots i_{m}}}\right)_{,i_{0}\cdots i_{j}}\\
=\sum_{h=1}^{p}\sum_{j=0}^{2p+2}\sum_{m=j+1}^{2p+3}\left(-1\right)^{j}\left(\sqrt{-g}a_{h}R\frac{\partial\Box^{h}R}{\partial g_{\mu\nu,\eta i_{0}\cdots i_{m}}}\right)_{,i_{0}\cdots i_{j}}\ .
\end{multline}
Thereby, after observing that $j+1\leq m \leq 2h+1$ $\rightarrow$ $j\leq 2h$, we finally get
\begin{equation*}
\sum_{j=0}^{n-2}\sum_{m=j+1}^{n-1}\left(-1\right)^{j}\left(\frac{\partial L}{\partial g_{\mu\nu,\eta i_{0}\cdots i_{m}}}\right)_{,i_{0}\cdots i_{j}}=\sum_{h=1}^{p}\sum_{j=0}^{2h}\sum_{m=j+1}^{2h+1}\left(-1\right)^{j}\left(\sqrt{-g}a_{h}R\frac{\partial\Box^{h}R}{\partial g_{\mu\nu,\eta i_{0}\cdots i_{m}}}\right)_{,i_{0}\cdots i_{j}}\ .
\end{equation*}
By inserting these expressions into \eqref{tensemN}, we obtain the gravitational energy--momentum pseudo-tensor for the Lagrangian \eqref{higher}
\begin{equation}\label{fulltensor}
{
\begin{split}
\tau_{\alpha}^{\eta}=\tau_{\alpha\vert GR}^{\eta}+&\frac{1}{2\chi\sqrt{-g}}\Biggl\{\sqrt{-g}\left(2a_{0}R+\sum_{k=1}^{p}a_{k}\Box^{k}R\right)\left[\frac{\partial R}{\partial g_{\mu\nu,\eta}}g_{\mu\nu,\alpha}+\frac{\partial R}{\partial g_{\mu\nu,\eta\lambda}}g_{\mu\nu,\lambda\alpha}\right]\\ 
&-\partial_{\lambda}\left[\sqrt{-g}\left(2a_{0}R+\sum_{k=1}^{p}a_{k}\Box^{k}R\right)\frac{\partial R}{\partial g_{\mu\nu,\eta\lambda}}\right]g_{\mu\nu,\alpha}\\
&+\Theta_{\left[1,+\infty\right[}\left(p\right)\sum_{h=1}^{p}\Biggl\{\sum_{q=0}^{2h+1}\left(-1\right)^{q}\partial_{i_{0}\cdots i_{q}}\biggl[\sqrt{-g}a_{h}R\frac{\partial \Box^{h}R}{\partial g_{\mu\nu,\eta i_{0}\cdots i_{q}}}\biggl]g_{\mu\nu,\alpha}\\
&+\sum_{j=0}^{2h}\sum_{m=j+1}^{2h+1}\left(-1\right)^{j}\partial_{i_{0}\cdots i_{j}}\biggl[\sqrt{-g}a_{h}R\frac{\partial \Box^{h}R}{\partial g_{\mu\nu,\eta i_{0}\cdots i_{m}}}\biggl]g_{\mu\nu,i_{j+1}\cdots i_{m}\alpha}\Biggr\}\\
&-\delta_{\alpha}^{\eta}\left(a_{0}R^{2}+\sum_{k=1}^{p} a_{k}R\Box^{k}R\right)\sqrt{-g}\Biggr\}
\end{split}
},
\end{equation}
where the notation
$
\partial_{i_{0}}=\mathbb{I}
$ is the identity operator and $\tau_{\alpha\vert GR}^{\eta}$ indicates the energy--momentum pseudo-tensor of general relativity \cite{DI} defined as
\begin{equation}\label{tensoreGR}
\tau_{\alpha\vert GR}^{\eta}=\frac{1}{2\chi}\left(\frac{\partial \overline{R}}{\partial g_{\mu\nu,\eta}}g_{\mu\nu,\alpha}-\delta^{\eta}_{\alpha} \overline{R}\right)\ .
\end{equation}
Given that only $\overline{R}$ contributes to the field equations we can replace scalar density $\sqrt{-g}R$ with $\sqrt{-g}\overline{R}$,which is not a scalar density. This makes the gravitational pseudo-tensor easier to manipulate and for a straightforward generalization of results see in Ref. \cite{PML}.

An important extension of local Lagrangian \eqref{higher} to non-local Lagrangian is possible allowing $p \rightarrow \infty$. 
Let $D^{p}$ be a linear differential operator defined by
\begin{equation}\label{52}
D^{p}=\sum_{k=0}^{p}a_{k}\Box^{k}\,.
\end{equation}
If the weak or strong convergence is guaranteed under suitable assumptions for the coefficients $a_{k}$ (e.g. $\sum_{k=0}^{\infty}\vert a_{k} \vert<\infty$ ) and for the domain of the operator $ D^{p}$, we obtain the following non-local operator $F\left(\Box\right)$
\begin{equation}
\lim_{p\rightarrow\infty}\sum_{k=0}^{p}a_{k}\Box^{k}=F\left(\Box\right)\,
\end{equation}
and also our local action becomes non local, i.e.
\begin{equation}
I=\int_{\Omega}d^{4}x\left[\overline{R}+RF\left(\Box\right)R\right]\sqrt{-g}\,.
\end{equation}
Accordingly integral operator acts as 
\begin{equation}
\Phi\left(x\right)=\int_{\Omega}d^{4}yF\left(x-y\right)R\left(x\right)=F\left(\Box\right)R\left(x\right)\ .
\end{equation}
Let us carry out now the limit $n\rightarrow \infty$ for the energy--momentum pseudo-tensor of $n$-order Lagrangian \eqref{tensemN}, we may obtain the non-local pseudo-tensor, that is
\begin{equation}\label{limitinftens}
\lim_{n\rightarrow\infty}\tau_{\alpha}^{\eta}\left(x\right)=\overline{\tau}_{\alpha}^{\eta}\left(x\right)\ .
\end{equation}
Whereas $\tau_{\alpha}^{\eta}\left(x\right)$ transforms as an affine tensor, we could show that also its limit for $n\rightarrow\infty$, i.e., $\overline{\tau}_{\alpha}^{\eta}\left(x\right)$, is an affine tensor. For an linear transformation 
\begin{equation}
x^{\prime\mu}=\Lambda^{\mu}_{\nu}x^{\nu}\qquad \vert \Lambda \vert \neq 0
\end{equation}
the following affine pseudo-tensor changes as 
\begin{equation}\label{trasfafftens}
\tau^{\eta}_{\alpha}\left(x\right)=\Lambda^{-1\eta}_{\ \ \ \sigma}\Lambda^{\tau}_{\alpha}\tau^{\prime\sigma}_{\tau}\left(x^{\prime}\right)\,.
\end{equation}
Substituting \eqref{trasfafftens} in \eqref{limitinftens}, we have
\begin{equation}
\overline{\tau}_{\alpha}^{\eta}\left(x\right)=\lim_{n\rightarrow\infty}\Lambda^{-1\eta}_{\ \ \ \sigma}\Lambda^{\tau}_{\alpha}\tau^{\prime\sigma}_{\tau}\left(x^{\prime}\right)=\Lambda^{-1\eta}_{\ \ \ \sigma}\Lambda^{\tau}_{\alpha}\lim_{n\rightarrow\infty}\tau^{\prime\sigma}_{\tau}\left(x^{\prime}\right)=\Lambda^{-1\eta}_{\ \ \ \sigma}\Lambda^{\tau}_{\alpha}\overline{\tau}^{\prime\sigma}_{\tau}\left(x^{\prime}\right)
\end{equation}
which implies that $\overline{\tau}^{\sigma}_{\tau}\left(x\right)$ transforms as an affine object also in the limit $n\rightarrow\infty$.

\subsection{The weak-field limit of energy--momentum pseudo-tensor}\label{EMTLO}

The gravitational energy--momentum pseudo-tensor \eqref{fulltensor} related to Lagrangian \eqref{higher} in weak field approximation can be performed perturbing weakly spacetime metric around the Minkowski metric $\eta_{\mu\nu}$ as 
\begin{equation}\label{53}
g_{\mu\nu}=\eta_{\mu\nu}+h_{\mu\nu}\qquad\mbox{being}\quad |h_{\mu\nu}|\ll 1\ ,
\end{equation}
where $h=\eta^{\mu\nu}h_{\mu\nu}$ is the trace of perturbation. 
Thus, we expand the energy--momentum pseudo-tensor to lower order in $h$, namely, retaining terms up to $h^2$. Let's see what becomes the weakly perturbed pseudo-tensor \eqref{tensoreGR} in harmonic coordinates where $g^{\mu\nu}\Gamma^{\sigma}_{\mu\nu}=0$. The quadratic part of the Ricci scalar $\overline{R}$ yields
\begin{equation}\label{54}
\overline{R}=-g^{\mu\nu}\left(\Gamma^{\rho}_{\mu\sigma}\Gamma^{\sigma}_{\nu\rho}\right)\ ,
\end{equation}
that is 
\begin{equation}\label{55}
\overline{R}=-\frac{1}{4}g^{\mu\nu}g^{\sigma\lambda}g^{\rho\epsilon}\left(g_{\epsilon\mu,\sigma}+g_{\epsilon\sigma,\mu}-g_{\mu\sigma,\epsilon}\right)\left(g_{\lambda\nu,\rho}+g_{\lambda\rho,\nu}-g_{\nu\rho,\lambda}\right)\ .
\end{equation}
Keeping terms up to second order in $h^{2}$, we get
\begin{equation}\label{56}
\left(\frac{\partial\overline{R}}{\partial g_{\alpha\beta,\gamma}}\right)^{\left(1\right)}\left(g_{\alpha\beta,\delta}\right)^{\left(1\right)}\stackrel{h^{2}}=\left(\frac{1}{2}h^{\alpha\beta\ \gamma}_{\ \ ,}h_{\alpha\beta,\delta}-h^{\gamma\alpha\ \beta}_{\ \ ,}h_{\alpha\beta,\delta}\right)\ ,
\end{equation}
according to 
\begin{multline}\label{57}
\frac{\partial\overline{R}}{\partial g_{\alpha\beta,\gamma}}g_{\alpha\beta,\delta}=-\frac{1}{4}\biggl\{\left(g^{\mu\beta}g^{\sigma\alpha}g^{\epsilon\gamma}+g^{\mu\gamma}g^{\sigma\alpha}g^{\beta\epsilon}-g^{\mu\alpha}g^{\sigma\gamma}g^{\beta\epsilon}\right)\left(g_{\epsilon\mu,\sigma}+g_{\epsilon\sigma,\mu}-g_{\sigma\mu,\epsilon}\right)\\
+\left(g^{\beta\nu}g^{\gamma\lambda}g^{\rho\alpha}+g^{\gamma\nu}g^{\beta\lambda}g^{\rho\alpha}-g^{\alpha\lambda}g^{\beta\nu}g^{\rho\gamma}\right)\left(g_{\lambda\nu,\rho}+g_{\lambda\rho,\nu}-g_{\nu\rho,\lambda}\right)\biggr\}g_{\alpha\beta,\delta}\ ,
\end{multline}
 and also 
\begin{equation}\label{58}
\overline{R}^{\left(2\right)}=-\frac{1}{4}\left(h^{\sigma\lambda}_{\ \ ,\rho}h_{\lambda\sigma,}^{\ \ \ \rho}-2h^{\sigma\lambda}_{\ \ ,\rho}h^{\rho}_{\ \lambda,\sigma}\right)\ .
\end{equation}
Hence, when we put these terms into \eqref{tensoreGR} , the stress--energy pseudo-tensor in general relativity up to order $h^{2}$ takes the form
 \begin{equation}\label{59}
 \tau_{\alpha\vert GR}^{\eta}=\frac{1}{2\chi}\left[\frac{1}{2}h^{\mu\nu,\eta}h_{\mu\nu,\alpha}-h^{\eta\mu,\nu}h_{\mu\nu,\alpha}-\frac{1}{4}\delta_{\alpha}^{\eta}\left(h^{\sigma\lambda}_{\ \ ,\rho}h_{\lambda\sigma}^{\ \ ,\rho}-2h^{\sigma\lambda}_{\ \ ,\rho}h^{\rho}_{\ \lambda,\sigma}\right)\right]\ .
\end{equation} 
Now, we have to expand to second order in $h$ the corrections of the pseudo-tensor \eqref{fulltensor} due to extended gravity terms. To lower order in $h$ we consider the following expansions
\begin{multline}\label{60}
\left(\frac{\partial R}{\partial g_{\mu\nu,\eta\lambda}}\right)^{\left(0\right)}=\frac{1}{2}\left(g^{\mu\eta}g^{\nu\lambda}+g^{\mu\lambda}g^{\nu\eta}-2g^{\mu\nu}g^{\eta\lambda}\right)^{\left(0\right)}\\
=\frac{1}{2}\left(\eta^{\mu\eta}\eta^{\nu\lambda}+\eta^{\mu\lambda}\eta^{\nu\eta}-2\eta^{\mu\nu}\eta^{\eta\lambda}\right)\ ,
\end{multline}
\begin{equation}\label{61}
 \left(\frac{\partial R}{\partial g_{\mu\nu,\eta\lambda}}\right)^{\left(0\right)}\left(g_{\mu\nu,\lambda\alpha}\right)^{\left(1\right)}=\left(h^{\lambda\eta}_{\ \ ,\lambda\alpha}-h^{,\eta}_{\ \ \alpha}\right)=\left(h^{\lambda\eta}-\eta^{\eta\lambda}h\right)_{,\lambda\alpha}\stackrel{\text{h.g.}}{=}-\frac{1}{2}h^{,\eta}_{\ \ \alpha}\ ,
\end{equation}
\begin{equation}\label{62}
 \left(\frac{\partial R}{\partial g_{\mu\nu,\eta\lambda}}\right)^{\left(0\right)}\left(g_{\mu\nu,\alpha}\right)^{\left(1\right)}=\left(h^{\lambda\eta}-\eta^{\eta\lambda}h\right)_{,\alpha}\ ,
\end{equation}
\begin{multline}\label{derivatsupnonsimm}
\left(\frac{\partial\Box^{h}R}{\partial g_{\mu\nu,\eta i_{0}\cdots i_{m}}}\right)^{\left(0\right)}=\left(\frac{\partial\Box^{h}R}{\partial g_{\mu\nu,\eta i_{0}\cdots i_{q}}}\right)^{\left(0\right)}=\left(\frac{\partial\Box^{h}R}{\partial g_{\mu\nu,\eta i_{0}\cdots i_{2h+1}}}\right)^{\left(0\right)}\\=\eta^{i_{2}i_{3}}\cdots \eta^{i_{2h}i_{2h+1}}\left(\eta^{\mu i_{1}}\eta^{\nu\eta}-\eta^{\mu\nu}\eta^{\eta i_{1}}\right)+\cdots\ .
\end{multline}
Then, we take into account only the terms up to $h^{2}$ in harmonic gauge, as 
\begin{equation}\label{primoterm}
\left(2a_{0}R+\sum_{k=1}^{p}a_{k}\Box^{k}R\right)\frac{\partial R}{g_{\mu\nu,\eta\lambda}}g_{\mu\nu,\lambda\alpha}\stackrel{h^{2}}{\stackrel{\text{h.g.}}{=}}\frac{1}{4}\left(\sum_{k=0}^{p}a_{k}\Box^{k+1}h\right)h^{,\eta}_{\ \ \alpha}+\frac{1}{4}a_{0}h^{,\eta}_{\ \ \alpha}\Box h\ ,
\end{equation}
\begin{multline}\label{secondoterm}
 -\partial_{\lambda}\left[\sqrt{-g}\left(2a_{0}R+\sum_{k=1}^{p}a_{k}\Box^{k}R\right)\frac{\partial R}{\partial g_{\mu\nu,\eta\lambda}}\right]g_{\mu\nu,\alpha}\stackrel{h^{2}}{\stackrel{\text{h.g.}}{=}}a_{0}\Box h_{,\lambda}\left(h^{\lambda\eta}-\eta^{\eta\lambda}h\right)_{,\alpha}\\
 +\frac{1}{2}\sum_{k=1}^{p}a_{k}\Box^{k+1}h_{,\lambda}\left(h^{\lambda\eta}-\eta^{\lambda\eta}h\right)_{,\alpha}\ ,
\end{multline}
\begin{multline}\label{formuladermax}
 \sum_{h=1}^{p}\sum_{q=0}^{2h+1}\left(-1\right)^{q}\partial_{i_{0}\cdots i_{q}}\left[\sqrt{-g}a_{h}R\frac{\partial\Box^{h}R}{\partial g_{\mu\nu,\eta i_{0}\cdots i_{q}}}\right]g_{\mu\nu,\alpha}\\\stackrel{h^{2}}{\stackrel{\text{h.g.}}{=}}\frac{1}{2}\sum_{h=1}^{p}a_{h}\Box^{h+1}h_{,\lambda}\left(h^{\eta\lambda}-\eta^{\eta\lambda}h\right)_{,\alpha}+\left(A_{p}\right)_{\alpha}^{\eta}\ ,
\end{multline}
\begin{multline}\label{formuladerivatesup}
 \sum_{h=1}^{p}\sum_{j=0}^{2h}\sum_{m=j+1}^{2h+1}\left(-1\right)^{j}\partial_{i_{0}\cdots i_{j}}\left[\sqrt{-g}a_{h}R\frac{\partial\Box^{h}R}{\partial g_{\mu\nu,\eta i_{0}\cdots i_{m}}}\right]g_{\mu\nu,i_{j+1}\cdots i_{m}\alpha}\stackrel{h^{2}}{\stackrel{\text{h.g.}}{=}}\frac{1}{4}\sum_{h=1}^{p}a_{h}\Box h \Box^{h} h^{,\eta}_{\ \ \alpha}\\
 +\frac{1}{2}\sum_{h=0}^{1}\sum_{j=h}^{p-1+h}\sum_{m=j+1-h}^{p}\left(-1\right)^{h}a_{m}\Box^{m-j}\left(h^{\eta\lambda}-\eta^{\eta\lambda}h\right)_{,i_{h}\alpha}\Box^{j+1-h}h_{,\lambda}^{\ \ i_{h}}+\left(B_{p}\right)_{\alpha}^{\eta}\ .
\end{multline}
In Eqs.~\eqref{formuladermax},~\eqref{formuladerivatesup} and \eqref{derivatsupnonsimm}, we have disregarded the index permutations ($\mu\nu$) and $\left(\eta i_{1}\cdots i_{2h+1}\right)$ because $\left(A_{p}\right)_{\alpha}^{\eta}$ and $\left(B_{p}\right)_{\alpha}^{\eta}$ terms, averaged on a suitable spacetime region, vanish, according to Appendix~\eqref{A}. Hence we calculated only the term deriving from \eqref{derivordsupsimm} without considering the index permutations ($\mu\nu$) and $\left(\eta i_{1}\cdots i_{2h+1}\right)$. This because, taking into account terms obtained from permutations in $\left(A_{p}\right)_{\alpha}^{\eta}$ and $\left(B_{p}\right)_{\alpha}^{\eta}$, averaged on a suitable spacetime region, we obtain that are equal to zero as we will see below in Appendix~\ref{A}. This mathematical trick is essential to calculated the averaged gravitational energy--momentum pseudo-tensor and the power emitted by a source.

So, by inserting equalities \eqref{primoterm}, \eqref{secondoterm}, \eqref{formuladermax} and \eqref{formuladerivatesup} into \eqref{fulltensor}, we find the extra term of pseudo-tensor $\tau^{\eta}_{\alpha}$ to second order owing to extension of general relativity , that we call $\tilde{\tau}^{\eta}_{\alpha}$, that is 
\begin{multline}\label{total}
\tilde{\tau}_{\alpha}^{\eta}\stackrel{h^{2}}{=}\frac{1}{2\chi}\Biggl\{\frac{1}{4}\left(\sum_{k=0}^{p}a_{k}\Box^{k+1}h\right)h^{,\eta}_{\ \ \alpha}+\frac{1}{2}\sum_{t=0}^{p}a_{t}\Box^{t+1}h_{,\lambda}\left(h^{\eta\lambda}-\eta^{\eta\lambda}h\right)_{,\alpha}\\
+\frac{1}{2}\sum_{h=0}^{1}\sum_{j=h}^{p}\sum_{m=j}^{p}\left(-1\right)^{h}a_{m}\Box^{m-j}\left(h^{\eta\lambda}-\eta^{\eta\lambda}h\right)_{,\alpha i_{h}}\Box^{j+1-h}h_{,\lambda}^{\ \ i_{h}}\\
+\frac{1}{4}\sum_{l=0}^{p} a_{l}\Box^{l}\left(h^{,\eta}_{\ \ \alpha}-\Box h\delta_{\alpha}^{\eta}\right)\Box h+\Theta_{\left[1,+\infty\right[}\left(p\right)\left[\left(A_{p}\right)_{\alpha}^{\eta}+\left(B_{p}\right)_{\alpha}^{\eta}\right]\Biggr\}\ ,
\end{multline}
where conventions used are
\begin{equation*}
\left(\right)_{,\alpha i_{0}}=\left(\right)_{,\alpha} \qquad h_{,\lambda}^{\ \ i_{0}}=h_{,\lambda}\ .
\end{equation*}
In summary, we can split the gravitational energy--momentum pseudo-tensor in the general relativity part and in the Extended Gravity part, that is
\begin{equation}\label{63}
\tau_{\alpha}^{\eta}\stackrel{h^{2}}=\tau_{\alpha\vert GR}^{\eta}+\tilde{\tau}_{\alpha}^{\eta}\ .
\end{equation}
Now in the particular case when $p$ is equal to $0$ and $1$, extended corrections of the pseudo-tensor $\tilde{\tau}^{\eta}_{\alpha}$ was derived. Then, for $p=0$, that is, $L_{g}=\left(\overline{R}+a_{0}R^{2}\right)\sqrt{-g}$ as in the case discussed in \cite{PML}, we obtain
\begin{equation*}
\tau_{\alpha}^{\eta}\stackrel{h^{2}}=\tau_{\alpha\vert GR}^{\eta}+\tilde{\tau}_{\alpha}^{\eta}\ ,
\end{equation*}
with
\begin{equation}\label{64}
\tilde{\tau}_{\alpha}^{\eta}\stackrel{h^{2}}=\frac{a_{0}}{2\chi}\left(\frac{1}{2}h^{,\eta}_{\ \ \alpha}\Box h+ h^{\eta}_{\ \lambda,\alpha}\Box h^{,\lambda}-h_{,\alpha}\Box h^{,\eta}-\frac{1}{4}\left(\Box h\right)^{2}\delta_{\alpha}^{\eta}\right)\ .
\end{equation}
While for $p=1$, that is $L_{g}=\left(\overline{R}+a_{0}R^{2}+a_{1}R\Box R\right)\sqrt{-g}$, one has 
\begin{equation*}
\tau_{\alpha}^{\eta}\stackrel{h^{2}}=\tau_{\alpha\vert GR}^{\eta}+\tilde{\tau}_{\alpha}^{\eta}\ ,
\end{equation*}
where extended corrections to pseudo-tensor are 
\begin{multline}\label{65}
\tilde{\tau}_{\alpha}^{\eta}\stackrel{h^{2}}=\frac{1}{2\chi}\Biggl\{\frac{1}{4}\left(2a_{0}\Box h+a_{1}\Box^{2}h\right)h^{,\eta}_{\ \ \alpha}+\frac{1}{2	}\left(2a_{0}\Box h_{,\lambda}+a_{1}\Box^{2}h_{,\lambda}\right)\left(h^{\eta\lambda}-\eta^{\eta\lambda}h\right)_{,\alpha}\\
+\frac{1}{2}a_{1}\Box\left(h^{\eta\lambda}-\eta^{\eta\lambda}h\right)_{,\alpha}\Box h_{,\lambda}+\frac{1}{2}a_{1}\left(h^{\eta\lambda}-\eta^{\eta\lambda}h\right)_{,\alpha}\Box^{2}h_{,\lambda}-\frac{1}{2}a_{1}\left(h^{\eta\lambda}-\eta^{\eta\lambda}h\right)_{,\sigma\alpha}\Box h_{,\lambda}^{\ \ \sigma}\\
+\frac{1}{4}a_{1}\Box h^{,\eta}_{\ \ \alpha}\Box h-\frac{1}{4}\delta_{\alpha}^{\eta}\left[a_{0}\left(\Box h\right)+a_{1}\left(\Box^{2} h\right)\right]\Box h+\left(A_{1}\right)_{\alpha}^{\eta}+\left(B_{1}\right)_{\alpha}^{\eta}\Biggr\}\ .
\end{multline}
The iteration can be performed to every $p$ introducing new contributions into dynamics.
\section{Power emitted carried by a gravitational wave }\label{MVEMT}
We wish to calculate the power emitted in the form of gravitational waves by an isolated massive system considering the local conservation of the energy--momentum pseudo-tensor \eqref{17}.
\subsection{The average of the energy--momentum pseudo-tensor}
Let us now regard the wavelike solutions of the linearized field equations in vacuum associated with Lagrangian \eqref{higher}, for details see Ref.~\cite{CCC}. Gravitational waves solutions can be expressed as 
\begin{equation}
\label{wave}
h_{\mu\nu}\left(x\right)=\sum_{m=1}^{p+2}\int_{\Omega}\frac{d^{3}\mathbf{k}}{\left(2\pi\right)^{3}}\left(B_{m}\right)_{\mu\nu}\left(\mathbf{k}\right)e^{i\left(k_{m}\right)_{\alpha}x^{\alpha}}+c.c.\ ,
\end{equation}
where
\begin{equation}\label{66}
\left(B_{m}\right)_{\mu\nu}\left(\mathbf{k}\right)=
\begin{cases}
C_{\mu\nu}\left(\mathbf{k}\right)& \quad \text{for}\quad m=1 \\
\frac{1}{3}\left[\frac{\eta_{\mu\nu}}{2}+\frac{\left(k_{m}\right)_{\mu}\left(k_{m}\right)_{\nu}}{k_{\left(m\right)}^{2}}\right]\text{A}_{m}\left(\mathbf{k}\right)&\quad \text{for} \quad m\geq2
\end{cases}\ ,
\end{equation}
with $C_{\mu\nu}\left(\mathbf{k}\right)$ related to transverse-traceless polarization tensor typical of general relativity and $\text{A}_{m}\left(\mathbf{k}\right)$ the amplitude of wave at $\mathbf{k}$ fixed.
Here "c.c." stands for the complex conjugate.
The trace of tensor \eqref{66} is 
\begin{equation}\label{67}
\left(B_{m}\right)_{\lambda}^{\lambda}\left(\mathbf{k}\right)=
\begin{cases}
C_{\lambda}^{\lambda}\left(\mathbf{k}\right)&\quad \text{for}\quad m=1 \\
\text{A}_{m}\left(\mathbf{k}\right)&\quad \text{for} \quad m\geq2
\end{cases}\ ,
\end{equation}
and the $k_{m}^{\mu}=\left(\omega_{m}, \mathbf{k}\right)$ is the wave vector with $k_{m}^{2}=\omega_{m}^{2}-\vert \mathbf{k} \vert ^{2}=\text{M}^{2}$ where $k_{1}^{2}=0$ and $k_{m}^{2}\neq 0$ for $m\geq 2$.
Keeping $\mathbf{k}$ fixed, we derive the following relations
\begin{align}\label{68}
 h^{\ \ \eta}_{,\alpha}=&2Re\left\{\sum_{j=1}^{p+2}\left(-1\right)\left(k_{j}\right)_{\alpha}\left(k_{j}\right)^{\eta}A_{j}e^{i k_{j}x}\right\}\ ,\\
 \Box^{m}h_{,\lambda}=&2 Re\left\{\left(-1\right)^{m}i\sum_{j=1}^{p+2}\left(k_{j}\right)_{\lambda}\left(k_{j}^{2}\right)^{m}A_{j}e^{ik_{j}x}\right\} \ ,\\
 \Box^{q}\left(h^{\eta\lambda}-\eta^{\eta\lambda}h\right)_{,\alpha}=&2Re\left\{\left(-1\right)^{q}i\sum_{l=1}^{p+2}\left(k_{l}\right)_{\alpha}\left(k_{l}^{2}\right)^{q}\left[\left(B_{l}\right)^{\eta\lambda}-\eta^{\eta\lambda}\left(B_{l}\right)_{\rho}^{\rho}\right]e^{ik_{l}x}\right\}\ ,\\
 \Box^{m}h_{,\lambda}^{\ \ \sigma}=&2 Re\left\{\left(-1\right)^{m+1}\sum_{j=1}^{p+2}\left(k_{j}\right)_{\lambda}\left(k_{j}\right)^{\sigma}\left(k_{j}^{2}\right)^{m}A_{j}e^{ik_{j}x}\right\}\ , \\
 \Box^{q}\left(h^{\eta\lambda}-\eta^{\eta\lambda}h\right)_{,\sigma\alpha}=&2Re\left\{\left(-1\right)^{q+1}\sum_{l=1}^{p+2}\left(k_{l}\right)_{\sigma}\left(k_{l}\right)_{\alpha}\left(k_{l}^{2}\right)^{q}\left[\left(B_{l}\right)^{\eta\lambda}-\eta^{\eta\lambda}\left(B_{l}\right)_{\rho}^{\rho}\right]e^{ik_{l}x}\right\}\ , \\
 \Box^{n}h=&2Re\left\{\left(-1\right)^{n}\sum_{r=2}^{p+2}\left(k_{r}^{2}\right)^{n}A_{r}e^{ik_{r}x}\right\}\ .
\end{align}
Now, we choose a domain of the spacetime $\Omega$ such that $\vert \Omega \vert \gg \frac{1}{\vert k\vert}$ \cite{WE}. Then, we can perform the average of the gravitational energy--momentum pseudo-tensor $\tau_{\alpha}^{\eta}$ over our region and all integrals, including terms such as $e^{i\left(k_{i}-k_{j}\right)_{\alpha}x^{\alpha}}$, tend to zero, by means of following identities 
\begin{equation}\label{69}
Re\{f\}Re\{g\}=\frac{1}{2}Re\{fg\}+\frac{1}{2}Re\{f\bar{g}\}\ ,
\end{equation}
\begin{equation}\label{70}
 \left(k_{l}\right)_{\lambda}\left[\left(B_{l}\right)^{\eta\lambda}-\eta^{\eta\lambda}\left(B_{l}\right)_{\rho}^{\rho}\right]=-\frac{\left(k_{l}\right)^{\eta}}{2}A_{l}\ .
\end{equation}
In the harmonic gauge, after averaging and some algebraic manipulations, we find (see Appendix~\ref{A})
\begin{align}\label{medievarie}
\left\langle\Box^{m}h_{,\lambda}\Box^{q}\left(h^{\eta\lambda}-\eta^{\eta\lambda}h\right)_{,\alpha}\right\rangle=&\left(-1\right)^{m+q+1}\sum_{l=2}^{p+2}\left(k_{l}\right)_{\alpha}\left(k_{l}\right)^{\eta} \left(k_{l}^{2}\right)^{\left(m+q\right)}\vert A_{l}\vert^{2}\ , \nonumber\\
\left\langle\Box^{m}h_{,\lambda}^{\ \sigma}\Box^{q}\left(h^{\eta\lambda}-\eta^{\eta\lambda}h\right)_{,\sigma\alpha}\right\rangle=&\left(-1\right)^{m+q+1}\sum_{l=2}^{p+2}\left(k_{l}\right)_{\alpha}\left(k_{l}\right)^{\eta} \left(k_{l}^{2}\right)^{\left(m+q\right)+1}\vert A_{l}\vert^{2}\ ,\nonumber\\
\left\langle \Box^{q}h_{\ \alpha}^{,\eta}\Box^{m}h\right\rangle=&2\left(-1\right)^{m+q+1}\sum_{r=2}^{p+2}\left(k_{r}\right)_{\alpha}\left(k_{r}\right)^{\eta} \left(k_{r}^{2}\right)^{\left(m+q\right)}\vert A_{r}\vert^{2}\ ,\nonumber\\
\left\langle\Box^{m}h\Box h\right\rangle=&2\left(-1\right)^{m+1}\sum_{j=2}^{p+2}\left(k_{j}^{2}\right)^{m+1}\vert A_{j}\vert^{2}\ ,\nonumber\\
\langle\left(A_{p}\right)_{\alpha}^{\eta}\rangle=&\langle\left(B_{p}\right)_{\alpha}^{\eta}\rangle=0\ .
\end{align}
A set of polarization tensors forming a basis for the linearized solutions $h_{\mu\nu}$ is given in Appendix~\ref{B}.
According to equalities \eqref{medievarie}, we can calculate the average value of the energy--momentum pseudo-tensor as
\begin{multline}\label{MEMT}
\left\langle\tau_{\alpha}^{\eta}\right\rangle=\frac{1}{2\chi}\left[\left(k_{1}\right)^{\eta}\left(k_{1}\right)_{\alpha}\left(C^{\mu\nu}C_{\mu\nu}^{*}-\frac{1}{2}\vert C_{\lambda}^{\lambda}\vert^{2}\right)\right]\\
+\frac{1}{2\chi}\left[\left(-\frac{1}{6}\right)\sum_{j=2}^{p+2}\left(\left(k_{j}\right)^{\eta}\left(k_{j}\right)_{\alpha}-\frac{1}{2}k_{j}^{2}\delta_{\alpha}^{\eta}\right)\vert A_{j}\vert^{2}\right]\\
+\frac{1}{2\chi}\Biggl\{\Biggl[\sum_{l=0}^{p}\left(l+2\right)\left(-1\right)^{l}a_{l}\sum_{j=2}^{p+2}\left(k_{j}\right)^{\eta}\left(k_{j}\right)_{\alpha}\left(k_{j}^{2}\right)^{l+1}\vert A_{j}\vert^{2}\Biggr]\\
-\frac{1}{2}\sum_{l=0}^{p}\left(-1\right)^{l}a_{l}\sum_{j=2}^{p+2}\left(k_{j}^{2}\right)^{l+2}\vert A_{j}\vert^{2}\delta_{\alpha}^{\eta}\Biggr\}\ ,
\end{multline}
with gravitational coupling ${\chi=\frac{8\pi G}{c^{4}}}$.
In TT gauge for the first mode associated with $k_{1}$ and only in harmonic gauge for residual modes $k_{m}$, in the momentum space, it gets
\begin{equation}\label{71}
\begin{cases}
 \left(k_{1}\right)_{\mu}C^{\mu\nu}=0 \quad \land \quad C_{\lambda}^{\lambda}=0&\quad \text{if}\quad m=1\\
 \left(k_{m}\right)_{\mu}\left(B_{m}\right)^{\mu\nu}=\frac{1}{2}\left(B_{m}\right)_{\lambda}^{\lambda}k^{\nu}& \quad \text{if} \quad m\geq2
\end{cases}\ .
\end{equation}
We now explore a gravitational wave propagating in the $+z$-direction at $\mathbf{k}$ fixed, with 4-wave vector given by $k^{\mu}=\left(\omega,0,0,k_{z}\right)$ where $\omega_{1}^{2}=k_{z}^{2}$ if $k_{1}^{2}=0$ and $ k_{m}^{2}=m^{2}=\omega_{m}^{2}-k_{z}^{2}$ otherwise with $k_{z}>0$. Accordingly the averaged time-space tensorial component which can be seen as flux of gravitational energy along the $z$ axis through the surface that delimits our domain $\Omega$, reads
\begin{multline}\label{72}
 \left\langle\tau_{0}^{3}\right\rangle=\frac{c^{4}}{8\pi G}\omega_{1}^{2}\left(C_{11}^{2}+C_{12}^{2}\right)+\frac{c^{4}}{16\pi G}\Biggl[\left(-\frac{1}{6}\right)\sum_{j=2}^{p+2}\omega_{j}k_{z}\vert A_{j}\vert^{2}\\
 +\sum_{l=0}^{p}\left(l+2\right)\left(-1\right)^{l}a_{l}\sum_{j=2}^{p+2}\omega_{j}k_{z}m_{j}^{2\left(l+1\right)}\vert A_{j}\vert^{2}\Biggr]\ .
\end{multline}

Finally, we can calculate the emitted power per unit solid angle $\Omega$, radiated by the localized sources, in a direction $\hat{x}$ at $\mathbf{k}$ fixed. By choosing of the suitable gauge, for the local conservation of the energy--momentum pseudo-tensor \eqref{17}, the power is given by
\begin{equation}\label{73}
 \frac{dP}{d\Omega}=r^2\hat{x}^{i}\left\langle\tau_{0}^{i}\right\rangle\ .
\end{equation}
By ranging the index $p$ of the pseudo-tensor \eqref{72} over $\{0,1,2\}$, we obtain the following three cases 
\\
for \emph{p}=0
\begin{gather}\label{74}
 \left\langle\tau_{0}^{3}\right\rangle=\frac{c^4\omega_{1}^{2}}{8\pi G}\left[C_{11}^{2}+C_{12}^{2}\right]+\frac{c^{4}}{16\pi G}\biggl\{\left(-\frac{1}{6}\right)\omega_{2}\vert A_{2}\vert^{2}k_{z}+2a_{0}\omega_{2}m_{2}^{2}\vert A_{2}\vert^{2}k_{z}\biggr\}\ ,
\end{gather}
for \emph{p}=1
\begin{multline}\label{75}
 \left\langle\tau_{0}^{3}\right\rangle=\frac{c^4\omega_{1}^{2}}{8\pi G}\left[C_{11}^{2}+C_{12}^{2}\right]+\frac{c^{4}}{16\pi G}\biggl\{\left(-\frac{1}{6}\right)\left(\omega_{2}\vert A_{2}\vert^{2}+\omega_{3}\vert A_{3}\vert^{3}\right)k_{z}\\
 +2a_{0}\left[\left(\omega_{2}m_{2}^{2}\vert A_{2}\vert^{2}+
 \omega_{3}m_{3}^{2}\vert A_{3}\vert^{2}\vert^{2}\right)k_{z}\right] -3a_{1}\left[\left(\omega_{2}m_{2}^{4}\vert A_{2}\vert^{2}+\omega_{3}m_{3}^{4}\vert A_{3}\vert^{2}\right)k_{z}\right]\biggr\}\ ,
\end{multline}
and for \emph{p}=2
\begin{multline}\label{76}
 \left\langle\tau_{0}^{3}\right\rangle=\frac{c^4\omega_{1}^{2}}{8\pi G}\left[C_{11}^{2}+C_{12}^{2}\right]+\frac{c^{4}}{16\pi G}\biggl\{\left(-\frac{1}{6}\right)\left(\omega_{2}\vert A_{2}\vert^{2}+\omega_{3}\vert A_{3}\vert^{3}+\omega_{4}\vert A_{4}\vert^{2}\right)k_{z}\\+2a_{0}\left[\left(\omega_{2}m_{2}^{2}\vert A_{2}\vert^{2}+\omega_{3}m_{3}^{2}\vert A_{3}\vert^{2}+\omega_{4}m_{4}^{2}\vert A_{4}\vert^{2}\right)k_{z}\right]\\
 -3a_{1}\left[\left(\omega_{2}m_{2}^{4}\vert A_{2}\vert^{2}+\omega_{3}m_{3}^{4}\vert A_{3}\vert^{2}+\omega_{4}m_{4}^{4}\vert A_{4}\vert^{2}\right)k_{z}\right]\\
 +4a_{2}\left[\left(\omega_{2}m_{2}^{6}\vert A_{2}\vert^{2}+\omega_{3}m_{3}^{6}\vert A_{3}\vert^{2}+\omega_{4}m_{4}^{6}\vert A_{4}\vert^{2}\right)\right]\biggr\}\ ,
\end{multline}
where the gravitational coupling $\chi$ has been explicitly indicated. By formulas~\eqref{74},~\eqref{75} and \eqref{76} it is obvious that the first term comes out of general relativity and the corrections strongly depends on $p$. In any context where corrections to general relativity can be investigated, this approach could constitute a paradigm to search for higher order effects.

\section{Energy--Momentum Complex of $f({\cal R})$ gravity in Palatini approach. }\label{paltinisection}
	
\subsection{The gravitational pseudo-tensor of $f({\cal R})$ gravity in Palatini formulation.}
In Palatini approach the metric tensor $g_{\mu\nu}$ and the connection $\Gamma^\alpha_{\mu\nu}$ are independent, that means that we do not assume any relation between the metric and the connection, and Riemann and Ricci tensors are, in general, defined as 
	\begin{align}\label{78}
		{\cal R}_{\mu\nu}(\Gamma)=& \partial_\alpha\Gamma^\alpha_{\mu\nu}-\partial_\nu\Gamma^\alpha_{\mu\alpha}+\Gamma^\alpha_{\mu\nu}\,\Gamma^\sigma_{\alpha\sigma} -\Gamma^\alpha_{\nu\lambda} \, \Gamma^\lambda_{\mu\alpha},
		\\
		{\cal R}(g,\Gamma) =& {\cal R}_{\mu\nu}(\Gamma)\, g^{\mu \nu}.
	\end{align} 
So, the Palatini gravitational action of $f({\cal R})$ appears as \cite{ACCA}
	\begin{align} \label{actionf}
		{\cal S} = \frac{1}{2 \kappa^2} \int {\rm d}^4x \,\sqrt{-g}\, f({\cal R}),
	\end{align}
with the coupling $\kappa^2=8\pi G /c^4$ and $g$ the determinant of metric tensor $g_{\mu\nu}$. By varying the metric $g^{\mu\nu}$ and the connection $\Gamma^\alpha_{\mu\nu}$, for a general infinitesimal transformation coordinate $x^\mu$ it gets
	\begin{align}
		x^{\prime \mu} =& x^\mu + \delta x^\mu,
		\\
		g^{\prime \mu\nu}(x^\prime) =& g^{\mu\nu}(x) + \tilde{\delta}g^{\mu\nu} ,&
		g^{\prime \mu\nu}(x) =& g^{\mu\nu}(x) + \delta g^{\mu\nu},
		\\
		\Gamma^{\prime \alpha}_{\mu\nu}(x^\prime) =& \Gamma^\alpha_{\mu\nu}(x) + \tilde{\delta} \Gamma^\alpha_{\mu\nu},&
		\Gamma^{\prime\alpha}_{\mu\nu}(x) =& \Gamma^\alpha_{\mu\nu}(x) + \delta \Gamma^\alpha_{\mu\nu},
	\end{align}
where $\tilde{\delta}$ is the local variation and $\delta$ is the variation that keeps the coordinates $x$ fixed.
The variation of the gravitational action with respect to the metric $g^{\mu\nu}$ and the connection $\Gamma^{\alpha}_{\beta\gamma}$ yield
	\begin{multline}\label{79}
		\tilde{\delta} {\cal S} = \frac{1}{2\kappa^2} \int {\rm d}^4x \Bigg\{ \sqrt{-g}\biggl[\left(f_{\cal R} {\cal R}_{\mu\nu}-\frac{1}{2} g_{\mu\nu} f\right) \, \delta g^{\mu\nu} \\
		+ f_{\cal R} g^{\mu\nu} \, \delta {\cal R}_{\mu\nu}\biggl] +\partial_\mu \left(\sqrt{-g} f \, \delta x^\mu\right) \Bigg\},
	\end{multline}
	where $f_{\cal R}:={\rm d} f({\cal R})/{\rm d}{\cal R}$. According to the following Palatini identity
	\begin{align}\label{80}
		\delta {\cal R}_{\mu\nu} = \nabla_\alpha\left(\delta \Gamma^\alpha_{\mu\nu}\right)-\nabla_\nu\left(\delta\Gamma^\alpha_{\alpha\mu}\right).
	\end{align}
the action \eqref{79} takes the form
	\begin{multline} \label{action1}
		\tilde{\delta} {\cal S} = \frac{1}{2\kappa^2 } \int {\rm d}^4x \, \Bigg\{ \sqrt{-g}\left(f_{\cal R} {\cal R}_{\mu\nu}-\frac{1}{2} g_{\mu\nu} f\right) \, \delta g^{\mu\nu} \\
		+\delta \Gamma^\lambda_{\phantom{\lambda}\nu\mu}\Bigl[-\nabla_\lambda\left(\sqrt{-g}g^{\nu\mu}f_{\cal R}\right)+\nabla_\alpha\left(\sqrt{-g}g^{\mu\alpha} \delta^\nu_\lambda f_{\cal R}\right)\Bigr]
		 \\
		+
		\partial_\lambda\left[\sqrt{-g}f_{\cal R} \left(g^{\mu\nu} \delta^\lambda_\alpha-g^{\mu\lambda} \delta^{\nu}_\alpha \right) \delta \Gamma^\alpha_{\phantom{\alpha}\mu\nu} +\sqrt{-g} f \, \delta x^\lambda \right]
		\Bigg\}.
	\end{multline}
By the principle of least action or stationary action \eqref{actionf}, by imposing that the variation of metric and its derivatives vanish at the boundary, we obtain field equations for the metric tensor and the connection in vacuum, i.e.,
	\begin{align}\label{FieldEq1}
		f_{\cal R} {\cal R}_{(\mu\nu)}-\frac{1}{2} g_{\mu\nu} f =& 0,
		\\ \label{FieldEq2}
		\nabla_\lambda\left(\sqrt{-g}g^{\nu\mu}f_{\cal R}\right) =& 0.
	\end{align}
Given that we adopting an arbitrary non-compatible connection, the symmetric part of the Ricci tensor, ${\cal R}_{(\mu\nu)}$, enter in the Eq.~\eqref{FieldEq1} and then the Ricci tensor is non symmetric, that is
	\begin{equation}\label{81}
	{\cal R}_{\mu\nu}={\cal R}_{\nu\mu}+{\cal R}^{\lambda}_{\phantom{\lambda}\lambda\mu\nu}\ ,
	\end{equation}
being Riemann tensor ${\cal R}^{\sigma}_{\phantom{\sigma}\lambda\mu\nu}$ no longer antisymmetric on its first two indices, i.e., the term ${\cal R}^{\lambda}_{\phantom{\lambda}\lambda\mu\nu}$ does not vanishes. For a generic infinitesimal transformation, the metric tensor and the connection change as
	\begin{align}\label{82}
		x^{\prime \mu} =& \, x^\mu + \xi^\mu,
		\\
		g^{\prime \mu\nu}(x^\lambda) \simeq& \, g^{\mu\nu}(x^\lambda) - \xi^\lambda \frac{\partial g^{\mu\nu}}{\partial x^\lambda} + g^{\mu\alpha} \frac{\partial \xi^\nu}{\partial x^\alpha} + g^{\nu\alpha} \frac{\partial \xi^\mu}{\partial x^\alpha},
		\\
		\Gamma^{\prime \alpha}_{\phantom{\alpha}\mu\nu} (x^\lambda) \simeq& \, \Gamma^{\alpha}_{\phantom{\alpha}\mu\nu} (x^\lambda) - \xi^\lambda \frac{\partial \Gamma^{\alpha}_{\phantom{\alpha}\mu\nu}}{\partial x^\lambda} + \Gamma^{\rho}_{\phantom{\rho}\mu\nu} \frac{\partial \xi^\alpha}{\partial x^\rho} -\Gamma^{\alpha}_{\phantom{\alpha}\sigma\nu} \frac{\partial \xi^\sigma}{\partial x^\mu} -\Gamma^{\alpha}_{\phantom{\alpha}\mu\sigma} \frac{\partial \xi^\sigma}{\partial x^\nu} - \frac{\partial^2 \xi^\alpha}{\partial x^\mu \, \partial x^\nu},
	\end{align}
	where we have neglected terms of higher order in $\xi^\mu$ in the series expansion.
	Under a rigid infinitesimal translation, that is, $\partial_\mu \xi^\nu =0$, we obtain
	\begin{align}
		g^{\prime \mu\nu}(x^\lambda) \simeq \, & g^{\mu\nu}(x^\lambda) - \xi^\lambda \frac{\partial g^{\mu\nu}}{\partial x^\lambda},
		\\
		\Gamma^{\prime \alpha}_{\phantom{\alpha}\mu\nu} (x^\lambda) \simeq \, & \Gamma^{\alpha}_{\phantom{\alpha}\mu\nu} (x^\lambda) - \xi^\lambda \frac{\partial \Gamma^{\alpha}_{\phantom{\alpha}\mu\nu}}{\partial x^\lambda} .
	\end{align}
Therefore, the Palatini action \eqref{action1} becomes
	\begin{multline}\label{83}
		\tilde{\delta} {\cal S}_{\rm g} = \frac{1}{2\kappa^2} \int {\rm d}^4x \Bigg\{ - \sqrt{-g}\left(f_{\cal R} {\cal R}_{\mu\nu}-\frac{1}{2} g_{\mu\nu} f\right) \, \xi^\lambda \, g^{\mu\nu}_{\phantom{\mu\nu},\lambda}\\
		-\xi^\lambda \, \Gamma^\beta_{\phantom{\beta}\nu\mu,\lambda}\left[-\nabla_\beta\left(\sqrt{-g}g^{\nu\mu}f_{\cal R}\right)+\nabla_\alpha\left(\sqrt{-g}g^{\mu\alpha} \delta^\nu_\beta f_{\cal R}\right)\right]
		 \\
		+
		\partial_\lambda\left[-\sqrt{-g}f_{\cal R} \left(g^{\mu\nu} \delta^\lambda_\alpha-g^{\mu\lambda} \delta^\nu_\alpha \right) \, \xi^\beta \, \Gamma^\alpha_{\phantom{\alpha}\mu\nu,\beta} +\sqrt{-g} f \, \xi^\lambda \right]
		\Bigg\}.
	\end{multline}
If the metric $g^{\mu\nu}$ and the Palatini connection $\Gamma^{\alpha}_{\beta\gamma}$ are solution of equations \eqref{FieldEq1} and \eqref{FieldEq2}, the stationary of the local variation of the action \eqref{83}, gives the local conservation of gravitational energy--momentum pseudo-tensor $\tau^\lambda_{\phantom{\lambda}\beta}$ of Palatini $f({\cal R})$ gravity, namely
	\begin{align}\label{84}
		\partial_\lambda\left(\sqrt{-g} \, \tau^\lambda_{\phantom{\lambda}\beta} \right) = 0,
	\end{align}
	where $\tau^\lambda_{\phantom{\lambda}\beta}$ is defined as
	\begin{align}\label{85}
		\tau^\lambda_{\phantom{\lambda}\beta}=\frac{1}{2\kappa^{2}} \left[f\left({\cal R}\right) \, \delta^\lambda_\beta -f_{\cal R}\left({\cal R}\right)\, \left(g^{\mu\nu} \delta^\lambda_\alpha-g^{\mu\lambda} \delta^\nu_\alpha \right) \, \Gamma^\alpha_{\phantom{\alpha}\mu\nu,\beta} \right].
	\end{align}
It is worth noting that the pseudo-tensor defined in Eq. \eqref{85} has the opposite sign of the one defined above. In order to derive the energy--momentum complex, let us analyze the action containing the matter part, that is 
	\begin{align}\label{86}
		{\cal S}_{\rm m} = \int {\rm d}^4x \, \sqrt{-g}\,{\cal L}_{\rm m}.
	\end{align}
Generally, the matter Lagrangian ${\cal L}_{\rm m}$ depends on the connection as, for example, occurs in presence of fermion fields.
	Here, we consider only material Lagrangian which does not depend on the affine connection $\Gamma$. Then, the matter energy--momentum tensor is defined as in \eqref{43}. 
Hence, field equations for metric and connection, i.e., Eqs. \eqref{FieldEq1} and \eqref{FieldEq2}, in presence of matter yield
	\begin{align}\label{Einstein and Palatini Eqs.}
		f_{\cal R} {\cal R}_{(\mu\nu)}-\frac{1}{2} g_{\mu\nu} f =& \kappa^2 T_{\mu\nu},
		\\
		\nabla_\lambda\left(\sqrt{-g}g^{\nu\mu}f_{\cal R}\right) =& 0 \label{bimetric}.
	\end{align}
As already pointed out above the connection can be non compatible with the metric $g_{\mu\nu}$, i.e., $\nabla_{\lambda}g_{\mu\nu}\neq 0$. In compact form, we can define a new metric, conformally related to the metric $g_{\mu\nu}$, as
	\begin{align}\label{87}
		h_{\mu\nu}:=f_{\cal R}g_{\mu\nu}.
	\end{align}
so that Eq. \eqref{bimetric} becomes 
	\begin{align}\label{88}
		\nabla_\lambda \left(\sqrt{h}h^{\mu\nu}\right)=0.
	\end{align}
	Thus the Palatini connection $\Gamma^\alpha_{\phantom{\alpha}\mu\nu} $ appears as the Christoffel connection for the new metric $h_{\mu\nu}$, i.e., \begin{align}\label{PalatiniConnection}
		\Gamma^\alpha_{\phantom{\alpha}\mu\nu}=
		\frac{1}{2\, f_{\cal R}\left({\cal R}\right)}g^{\alpha \beta} \left[\partial_\mu\left( f_{\cal R}\left({\cal R}\right) g_{\nu\beta}\right)+\partial_\nu\left( f_{\cal R}\left({\cal R}\right) g_{\mu\beta}\right)-\partial_\beta\left( f_{\cal R}\left({\cal R}\right) g_{\mu\nu}\right)\right]\ .
	\end{align}
	The Palatini connection $\Gamma^\alpha_{\phantom{\alpha}\mu\nu} $ and Levi--Civita connection $\stackrel{\circ}{\Gamma}{}^\alpha_{\phantom{\alpha}\mu\nu} $ are related as 
	\begin{equation}\label{89}
	\Gamma^\alpha_{\phantom{\alpha}\mu\nu}=\,\stackrel{\circ}{\Gamma^{\alpha}}_{\mu\nu}+\delta^{\alpha}_{\mu}A_{\nu}+\delta^{\alpha}_{\nu}A_{\mu}-g_{\mu\nu}A^{\alpha}\ ,
	\end{equation}\label{90}
	where the four-vector $A_{\mu}$ is defined as
	\begin{equation}\label{91}
	A_{\mu}:=\frac{1}{2f_{\cal R}}\nabla_{\mu}f_{\cal R}\ .
	\end{equation}
	For $f({\cal R})={\cal R}$, we recover the Christoffel symbols constructed by the metric $g_{\mu\nu}$, that is 
	\begin{align}\label{92}
	\Gamma^\alpha_{\phantom{\alpha}\mu\nu}=\stackrel{\circ}{\Gamma^\alpha}_{\phantom{\alpha}\mu\nu} = \frac{1}{2}g^{\alpha\beta}\left( g_{\beta\mu,\nu}+g_{\beta\nu,\mu}-g_{\mu\nu,\beta}\right)\,
	\end{align}
this means that in general relativity no difference results in metric and Palatini formalism. The Ricci tensor ${\cal R}_{\mu\nu}$ in Palatini formalism and that in metric formalism $R_{\mu\nu}$, are related as follows
	\begin{multline} \label{Rconf}
		{\cal R}_{\mu\nu} = R_{\mu\nu} +\frac{3}{2} \frac{1}{\left(f_{\cal R}({\cal R})\right)^2} \left(\stackrel{\circ}{\nabla}_\mu f_{\cal R}({\cal R}) \right) \, \left(\stackrel{\circ}{\nabla}_\mu f_{\cal R}({\cal R})\right) \\
		- \frac{1}{ f_{\cal R}({\cal R})} \left( \stackrel{\circ}{\nabla}_\mu \stackrel{\circ}{\nabla}_\nu-\frac{1}{2}g_{\mu\nu} \stackrel{\circ}{\square}\right) f_{\cal R}({\cal R}),
	\end{multline}
	where $\stackrel{\circ}{\square} := \stackrel{\circ}{\nabla}{}^\mu \stackrel{\circ}{\nabla}_\mu$ and $\stackrel{\circ}{\nabla}$ denotes the covariant derivative associated with the Levi--Civita connection. Contracting tensorial equality \eqref{Rconf} with $g^{\mu\nu}$, we obtain the relation between ${\cal R}$ and $R$, that is, the Ricci scalar in both approach
	\begin{align}\label{93}
		{\cal R} = R + \frac{3}{2\left(f_{\cal R}({\cal R})\right)^2} \left(\stackrel{\circ}{\nabla}_\mu f_{\cal R}({\cal R}) \right)\; \left(\stackrel{\circ}{\nabla}{ }^\mu f_{\cal R}({\cal R})\right) + \frac{3}{ f_{\cal R}({\cal R})} \stackrel{\circ}{\square} f_{\cal R}({\cal R}).
	\end{align}
Adopting the Palatini connection $\Gamma^\alpha_{\phantom{\alpha}\mu\nu}$ \eqref{PalatiniConnection}, the symmetry of Ricci tensor is restored on account of the relation
	\begin{equation}\label{94}
	\Gamma_{\lambda}=\frac{\partial_{\lambda}{\left(f_{\cal R}^{2}\sqrt{-g}\right)}}{f_{\cal R}^{2}\sqrt{-g}},
	\end{equation}
	which implies
	\begin{equation}\label{95}
	{\cal R}_{[\mu\nu]}=\partial_{[\mu}\Gamma_{\nu]}=0.
	\end{equation}
Furthermore the connection is non compatible with metric $g_{\mu\nu}$ being
	\begin{equation}\label{96}
	\nabla_{\lambda}g_{\mu\nu}=-\frac{g_{\mu\nu}}{f_{\cal R}}\nabla_{\lambda}f_{\cal R}.
	\end{equation}
	Despite this, the covariant derivatives associated with Palatini connection commute each other, as displayed below
	\begin{equation}\label{97}
	\left[\nabla_{\rho},\nabla_{\lambda}\right]g_{\mu\nu}=0\,.
	\end{equation}
	Thus, we restore the antisymmetry on the first two indices of Riemann tensor, namely
	\begin{equation}\label{99}
	{\cal R}_{\mu\nu\lambda\rho}=-{\cal R}_{\nu\mu\lambda\rho}.
	\end{equation}
 by the definition of Riemann tensor for an arbitrary tensor $J_{\mu\nu}$
	\begin{equation}\label{98}
	\left[\nabla_{\rho},\nabla_{\lambda}\right]J_{\mu\nu}=-{\cal R}^{\alpha}_{\phantom{\alpha}\mu\rho\lambda}J_{\alpha\nu}-{\cal R}^{\alpha}_{\phantom{\alpha}\nu\rho\lambda}J_{\mu\alpha}.
	\end{equation}
	In addition, the contracted Bianchi identities are fulfilled, that is 
	\begin{equation}\label{100}
	\nabla_{\mu}\left({\cal R}^{\mu\nu}-\frac{1}{2}g^{\mu\nu}{\cal R}\right)=0.
	\end{equation}
According to the Palatini connection Eq. \eqref{PalatiniConnection} and from the symmetry of energy--momentum tensor $T_{\mu\nu}$, taking into account that for the new metric $h_{\mu\nu}$ we have
	\begin{align}\label{101}
		\Gamma_{\lambda}=\frac{\partial_{\lambda}{\sqrt{-h}}}{\sqrt{-h}}\ ,
	\end{align}
	and 
	\begin{align}\label{102}
		\Gamma_{\mu\nu\lambda}+\Gamma_{\nu\mu\lambda}=\frac{1}{f_{{\cal R}}}\partial_{\lambda}h_{\mu\nu},
	\end{align}
so we derive the following useful expression 
	\begin{align}\label{103}
		\sqrt{-h}\nabla_{\sigma}T^{\sigma}_{\phantom{\sigma}\nu}=\partial_{\sigma}\left(\sqrt{-h}T^{\sigma}_{\phantom{\sigma}\nu}\right)-\frac{1}{2f_{\cal R}}T^{\lambda\rho}\partial_{\nu}h_{\lambda\rho}\sqrt{-h}.
	\end{align}
Field equations in matter \eqref{Einstein and Palatini Eqs.} lead to 
	\begin{align}\label{104}
		0=\frac{\sqrt{-h}}{2f_{\cal R}^{2}}T^{\mu\nu}g_{\mu\nu,\beta}\xi^{\beta}+\partial_{\lambda}\left\{\sqrt{-g}\frac{1}{2\kappa^{2}}\left[
		f\left({\cal R}\right) \, \delta^\lambda_\beta -f_{\cal R}\left(g^{\mu\nu} \delta^\lambda_\alpha-g^{\mu\lambda} \delta^\nu_\alpha \right) \, \Gamma^\alpha_{\phantom{\alpha}\mu\nu,\beta} \right]\xi^{\beta}\right\},
	\end{align}
and from Eq. \eqref{103}, after some algebraic manipulations, we get the following 4-divergence of energy--momentum complex not vanishing
	\begin{align}\label{105}
		\partial_{\sigma}\left[\sqrt{-g}\left(T^{\sigma}_{\phantom{\sigma}\beta}+t^{\sigma}_{\phantom{\sigma}\beta}\right)\right]=\frac{\sqrt{-h}}{f_{\cal R}^2} \, \nabla_{\lambda}T^{\lambda}_{\phantom{\lambda}\beta}+\frac{2\sqrt{-h}}{f_{\cal R}^{3}}T^{\lambda}_{\phantom{\lambda}\beta} \, \nabla_{\lambda}f_{\cal R}-\frac{\sqrt{-h}}{2f_{\cal R}^{3}}T \, \nabla_{\beta}f_{\cal R}.
	\end{align}
	From contracted Bianchi identities and the field equations, the following relations are satisfied
	\begin{equation}\label{106}
	\left[\nabla_{\mu},\nabla_{\nu}\right]\nabla^{\mu}f_{\cal R}={\cal R}^{\alpha}_{\phantom{\alpha}\nu}\nabla_{\alpha}f_{\cal R},
	\end{equation}
	and
	\begin{equation}\label{107}
	\kappa^{2}\nabla_{\mu}T^{\mu}_{\phantom{\mu}\nu}={\cal R}^{\alpha}_{\phantom{\alpha}\nu}\nabla_{\alpha}f_{\cal R}=\left[\nabla_{\mu},\nabla_{\nu}\right]\nabla^{\mu}f_{\cal R}\ .
	\end{equation}
	The trace of Eqs. \eqref{Einstein and Palatini Eqs.} gives the so called {\it structural equation} of space-time \cite{Mauro}, that is 
	\begin{equation}\label{108}
	T=\frac{1}{\kappa^{2}}\left[f_{\cal R}{\cal R}-2f\left({\cal R}\right)\right],
	\end{equation}
	where $T=T_{\mu\nu}g^{\mu\nu}$. For a given $f({\cal R})$, we can, in principle, solve this equation and get a relation ${\cal R}={\cal R}(T)$. Thanks to Eq. \eqref{108}, considering $T=0$, the theory reduces to GR with a cosmological constant. Substituting Eqs. \eqref{107} and \eqref{108} into Eq. \eqref{105}, we get 
	\begin{equation} \label{109}
	\partial_{\sigma}\left[\sqrt{-g}\left(T^{\sigma}_{\phantom{\sigma}\beta}+\tau^{\sigma}_{\phantom{\sigma}\beta}\right)\right]=-\frac{\sqrt{-g}}{\kappa^{2}}G^{\lambda}_{\phantom{\lambda}\beta}\nabla_{\lambda}f_{\cal R},
	\end{equation}
	where $G^{\lambda}_{\phantom{\lambda}\beta}$ is the Einstein tensor.
	After some algebraic manipulations, we find the following expression 
	\begin{equation}\label{110}
	G^{\lambda}_{\phantom{\lambda}\beta}\nabla_{\lambda}f_{\cal R}=-\kappa^{2}\stackrel{\circ}{\nabla}_{\mu}T^{\mu}_{\phantom{\mu}\beta}\ .
	\end{equation}
	The right hand side of Eq. \eqref{110} vanishes \cite{Dick:1992jn, Barraco:1998eq, Koivisto:2005yk} and then, according to Eqs. \eqref{109} and \eqref{110}, the energy--momentum complex for Palatini $f({\cal R})$ gravity, $\mathcal{T}^{\sigma}_{\phantom{\sigma}\beta}$, is locally conserved, namely
	\begin{equation}
	\partial_{\sigma}\left[\sqrt{-g}\left(T^{\sigma}_{\phantom{\sigma}\beta}+\tau^{\sigma}_{\phantom{\sigma}\beta}\right)\right]=0,
	\end{equation}
with
\begin{equation}\label{111}
\mathcal{T}^{\sigma}_{\phantom{\sigma}\beta}=\sqrt{-g}\left(T^{\sigma}_{\phantom{\sigma}\beta}+\tau^{\sigma}_{\phantom{\sigma}\beta}\right)\ .
\end{equation}

\section{Cosmological applications both in Palatini and metric approach in $f(R)$ gravity}\label{cosmo}
\subsection{Palatini formalism}	

We consider a flat FLRW spacetime whose metric is 
	\begin{align}
		{\rm d}s^2= -{\rm d}t^2+ a^2(t)\, \left({\rm d}x^2+{\rm d}y^2+{\rm d}z^2\right),
	\end{align}
	with scale factor $a(t)$ and cosmic time $t$. From the relation \eqref{Rconf} and the field equations \eqref{Einstein and Palatini Eqs.}, we obtain
	\begin{align}
		2\kappa^2 T^0_{\phantom{0}0} = & -f +6f_{\cal R} \left(\dot{H}+H^2\right) +\ddot{f}_{\cal R} -3\frac{\dot{f}_{\cal R}^2}{f_{\cal R}} -3H\dot{f}_{\cal R},
	\end{align}
	where $H=\dot{a}/a$ is the Hubble parameter and dots stands for derivatives with respect to the cosmic time $t$. The gravitational energy density $\tau^0_{\phantom{0}0}$ is defined as
	\begin{align}
		2\kappa^2 \tau^0_{\phantom{0}0} = & f -6f_{\cal R} \left(\dot{H}+H^2\right) -3\ddot{f}_{\cal R} +3\frac{\dot{f}_{\cal R}^2}{f_{\cal R}} -3H\dot{f}_{\cal R}.
	\end{align}
	So, the energy density complex is 
	\begin{align}
		\kappa^2(\tau^0_{\phantom{0}0}+T^0_{\phantom{0}0}) = -\ddot{f}_{\cal R} -3H \dot{f}_{\cal R}.
	\end{align} 
	In general relativity , i.e., $f({\cal R})={\cal R}$, we obtain a null energy density complex
	\begin{align}
	\left(t^0_{\phantom{0}0}+T^0_{\phantom{0}0}\right)\Big|_{\rm GR}=0\,.
	\end{align}
	We postulate that perfect fluids including radiation and non-relativistic dust describe the matter and that the components of the energy--momentum tensor are
	\begin{align}
		\left(T^\mu_{\phantom{\mu}\nu} \right)_{\rm r} =& {\rm diag}\left(-\rho_{\rm r},p_{\rm r},p_{\rm r},p_{\rm r} \right) & \mbox{ with equation of state}& &p_{\rm r}=&\frac{1}{3} \rho_{\rm r},
		\\
		\left(T^\mu_{\phantom{\mu}\nu} \right)_{\rm m} =& {\rm diag}\left(-\rho_{\rm m},p_{\rm m},p_{\rm m},p_{\rm m} \right), & \mbox{ with equation of state }& &p_{\rm m}=&0,
	\end{align}
	where $\rho_{\rm i}$ and $p_{\rm i}$ are the energy density and pressure of each fluid component.
	From the conservation of energy--momentum tensor, we obtain, respectively, 
	\begin{align} \label{c1}
		\dot{\rho}_{\rm r}+4 H \rho_{\rm r}=&0,
		\\ \label{c2}
		\dot{\rho}_{\rm m}+3 H \rho_{\rm m}=&0.
	\end{align}
	Choosing a form for $f({\cal R})$, we can solve the structure equation Eq. \eqref{108} and then explicit ${\cal R}$ as a function of $T$. Now, let us assume a polynomial form as $f({\cal R}) = {\cal R}+\alpha {\cal R}^2$, which is a model extensively studied in Palatini formalism, see for example \cite{Barragan:2009sq, Szydlowski:2015fcq}. Thus, the solution of structural equation \eqref{108} becomes
	\begin{align}
		{\cal R} = -\kappa^2 T.
	\end{align}
This model implies power law cosmological solutions \cite{Goheer:2009ss} as
	\begin{align} \label{powerlaw}
		a(t) = a_0 \, t^m,
	\end{align}
	where $m>0$ is a real number. From Eqs. \eqref{c1} and \eqref{c2}, we get
	\begin{align}
		\rho_{\rm tot}(t) = \rho_{\rm m}(t)+\rho_{\rm r}(t)=\rho_{\rm m0}t^{-3m}+\rho_{\rm r0}t^{-4m}\,
	\end{align}
	with $\rho_{\rm m0}$ and $\rho_{\rm r0}$ initial values.
	Therefore, we obtain the gravitational energy density
	\begin{multline}
		2 \kappa^2 \, \tau^0_{\phantom{0}0} = \frac{6m(1-m)}{t^2} +\kappa^2 \rho_{\rm m0}t^{-3m} +6m (5-2m) \alpha \kappa^2 \rho_{\rm m0} t^{-3m-2} \\
		+\frac{108m^2 \alpha^2 \kappa^4 \rho_{\rm m0}t^{-6m-2}}{1+2m^2\kappa^2\rho_{\rm m0} t^{-3m}},
	\end{multline}
	and the energy density complex 
	\begin{align}
	\tau^0_{\phantom{0}0}+T^0_{\phantom{0}0} =-6\alpha m \rho_{\rm m0} t^{-(3m+2)}.
	\end{align}
	The total energy density of gravitational and non-gravitational fields is then
	\begin{equation}\label{300}
	\sqrt{-g}(\tau^0_{\phantom{0}0}+T^0_{\phantom{0}0})=-6\alpha m \rho_{\rm m0} t^{-2},
	\end{equation}
	that tends to zero as the inverse square of cosmic time.

\subsection{Metric approach}

We consider also in this case a flat FLRW spacetime but in metric formalism. We can explicitly write the time-time components of the gravitational energy--momentum $\tau^\mu_{\phantom{\mu}\nu}$ and the matter energy--momentum $T^\mu_{\phantom{\mu}\nu}$, respectively, as follows
	\begin{align}
		\kappa^2\,	\tau^0_{\phantom{0}0} =& \frac{1}{2}f(R)-3\left(H^2+\dot{H}\right)\, f_R(R) +3H\dot{R} \, f_{RR}(R),
		\\ \label{f1}
		\kappa^2 T^0_{\phantom{0}0} =& - \frac{1}{2}f(R)+3\left(H^2+\dot{H}\right)\, f_R(R) - 3H\dot{R} \, f_{RR}(R).
	\end{align}
	Subsequently, the total energy of the gravitation and matter vanishes for FLRW spacetime, i.e.
	\begin{align}
	\tau^0_{\phantom{0}0}+T^0_{\phantom{0}0} =0,
	\end{align} 
unlike Palatini approach where energy complex does not vanish Eq.\eqref{300}. Now, we can assume a power-law evolution for matter and radiation fluids such as Eq. \eqref{powerlaw}. We have
	 \begin{align} \tau^0_{\phantom{0}0} =&\rho_{\rm m}(t)+\rho_{\rm r}(t)\nonumber \\ =&\rho_{\rm m0} t^{-3m}+\rho_{\rm r0}t^{-4m}\,.
	 \end{align}
 The Ricci curvature scalar, in this case, reads
	\begin{equation}
		R = 12H^2+6\dot{H} = 6m(2m-1) t^{-2}\,
	\end{equation}
while the Friedman equation is reduced to
	\begin{multline}\label{200}
		\frac{f_{RR}\, R^2}{(2m-1)}+\frac{m-1}{2(2m-1)} f_R \, R -\frac{1}{2} f +\kappa^2 \rho_{\rm m0} \left( \frac{R}{6m(2m-1)}\right)^{\frac{3}{2}m}\\
		+\kappa^2 \rho_{\rm r0} \left( \frac{R}{6m(2m-1)}\right)^{2m} =0.
	\end{multline}
	From this equation \eqref{200}, we get the explicit form of $f(R)$ that shows a power law behaviour, that is
	\begin{align} \label{f}
		f(R) =& -\frac{4\kappa^2 \rho_{\rm m0} (2m-1)}{12m-11} \left( \frac{R}{6m(2m-1)}\right)^{\frac{3}{2}m} -\frac{2\kappa^2 \rho_{\rm r0} (2m-1)}{10m^2-8m+1} \left( \frac{R}{6m(2m-1)}\right)^{2m} 
		\nonumber \\
		&+C_1 R^{\frac{3}{4}-\frac{m}{4}-\frac{1}{4}\sqrt{m^2+10m+1}} +C_2 R^{\frac{3}{4}-\frac{m}{4}+\frac{1}{4}\sqrt{m^2+10m+1}}.
	\end{align}
	When $m=2/3$ and $\rho_{\rm r0}/\rho_{\rm m0} \ll 1$, occurs $f(R)\sim R$ and GR is restored.

\section{Conclusions}\label{conclusions}

Attempts to extend the general relativity through corrections to the Hilbert--Einstein Lagrangian, by introducing curvature, torsion and non-metricity invariants, both local and non-local, have increased in recent years. All of this is to address gravitational divergences at ultraviolet and infrared scales, and more generally to deal with cosmological and astrophysical issues such as current and early cosmic acceleration or the structure formation, without introducing exotic components such as dark energy and dark matter. For a detailed discussion on infinite derivative theories, see Ref. \cite{BLP,BLMTY,BLY, BGLM,BCHKLM,BLM,BKLM}, while for non-local wavelike solutions, see Ref. \cite{CCCQG2021,CCN, CAPRIOLOM}. However, most of the main features of general relativity should be retained to obtain self-consistent theories. In particular, a thorough study of the properties of the gravitational energy--momentum pseudo-tensor are indispensable in view of both the foundation and applications of any gravitational theory.

This review is devoted to generalizing the gravitational energy--momentum pseudo-tensor $\tau^{\eta}_{\alpha}$ to general $n^{th}$ order Lagrangian of the form 
\begin{equation*}
L=L(g_{\mu\nu}, g_{\mu\nu,i_{1}}, g_{\mu\nu,i_{1}i_{2}},g_{\mu\nu,i_{1}i_{2}i_{3}}, \ldots, g_{\mu\nu,i_{1}i_{2}i_{3}\cdots i_{n}})\ ,
\end{equation*}
showing that in this model gravity a local conservation of energy--momentum complex is fulfilled.
Specifically, we considered Lagrangians such as $L_{g}=(\overline{R}+a_{0}R^{2}+\sum_{k=1}^{p} a_{k}R\Box^{k}R)\sqrt{-g}$ and $L=F(R)$, both in Palatini and metric approach. It has been shown that $\tau^{\eta}_{\alpha}$ is an affine and not covariant object because it changes as a tensor, under linear transformations but not under general coordinate transformations. The pseudo-tensor of higher order gravity has been weakly perturbed up to the order $h^2$, thus obtaining the weak field limit of the gravitational energy--momentum density. After averaging this object over a suitable four-dimensional domain under suitable gauge, by local conservation of pseudo-tensor, the power emitted by a gravitational source was found. Hence, the gravitational wave \eqref{wave} associated with higher order Lagrangian express, under the chosen gauge for a wave propagating along the $+z$-direction, in terms of six polarization tensors (see Appendix~\ref{B}) reads as 
\begin{multline}
\label{GW1}
h_{\mu\nu}\left(t;z\right)=\text{A}^{\left(+\right)}\left(t-z\right)\epsilon_{\mu\nu}^{\left(+\right)}
+\text{A}^{\left(\times\right)}\left(t-z\right)\epsilon_{\mu\nu}^{\left(\times\right)}+\text{A}^{\left(TT\right)}\left(t-v_{G_{m}}z\right)\epsilon_{\mu\nu}^{\left(TT\right)}\\
+\text{A}^{\left(TS\right)}\left(t-v_{G_{m}}z\right)\epsilon_{\mu\nu}^{\left(TS\right)}
+\text{A}^{\left(1\right)}\left(t-v_{G_{m}}z\right)\epsilon_{\mu\nu}^{\left(1\right)}+\text{A}^{\left(L\right)}\left(t-v_{G_{m}}z\right)\epsilon_{\mu\nu}^{\left(L\right)}\ ,
\end{multline}
where $v_{G_{m}}$ is the group velocity of the $m_{th}$ massive mode (see also \cite{greci, arturo1}). Thanks to these solutions, it was possible to derive an expression of the power emitted in terms of amplitudes of the waves $\text{A}_{j}\left(\mathbf{k}\right)$, $C_{11}\left(\mathbf{k}\right)$ and $C_{22}\left(\mathbf{k}\right)$, and the free parameters $a_{m}$. Three special cases for $p$ equal to $0,1,$ and $2$ have been shown where the extended corrections to the power are clearly visible. It was given a cosmological application of the pseudo-tensor in $f(R)$ gravity in both Palatini and metric formulation. Therefore, in a flat FLRW spacetime, we have derived that while the energy density complex vanishes in the metric formalism, in general, it does not vanish in the Palatini approach. 

The analysis of gravitational waves and gravitational energy--momentum pseudo-tensor 
are two indispensable tools for finding the viable theory of gravitation. Indeed, by wavelike solutions of linearized theory of gravity and by the locally conserved pseudo-tensor, it is possible to calculate the emitted power by isolated system. Then, from the local conservation of the energy--momentum complex, it is also possible to take into account the energy--momentum content of the source, which, through a multipole expansion, could also allow us to derive a generalized formula of the quadrupole formula. This procedure could lead us to fix the order of theory \cite{Quandt,staro}, to investigate additional polarization states of gravitational wave and to establish the range of the masses $m_{j}$ of modes.

\vspace{6pt}

\section*{Acknowledgments}
{S.C., M.C. and G.L. acknowledgment the Istituto Nazionale di Fisica Nucleare (INFN) Sez. di Napoli, Iniziative Specifiche QGSKY, and the Istituto Nazionale di Alta Matematica (INdAM), gruppo GNFM, for the support. }


\appendix
\section[\appendixname \thesection]{Appendix}
\subsection{The average of $\langle\left(A_{p}\right)_{\alpha}^{\eta}\rangle$ and $\langle\left(B_{p}\right)_{\alpha}^{\eta}\rangle$ terms}\label{A}

Let us now demonstrate the last two relations in \eqref{medievarie}, that is $\langle\left(A_{p}\right)_{\alpha}^{\eta}\rangle=\langle\left(B_{p}\right)_{\alpha}^{\eta}\rangle=0$.
The general formula for $\Box^{h} R$ -derivative, according to symmetries of $g_{\mu\nu}$ and its derivatives, is \cite{staro}:
\begin{multline}\label{derivordsupsimm}
\frac{\partial \Box^{h}R}{\partial g_{\mu\nu,\eta i_{1}\cdots i_{2h+1}}}=g^{j_{2}j_{3}}\cdots g^{j_{2h}j_{2h+1}}g^{ab}g^{cd}\biggl\{\delta_{a}^{(\mu}\delta_{d}^{\nu)}\delta_{c}^{(\eta}\delta_{b}^{i_{1}}\delta_{j_{2}}^{i_{2}}\cdots\delta_{j_{2h}}^{i_{2h}}\delta_{j_{2h+1}}^{i_{2h+1})}\\
-\delta_{a}^{(\mu}\delta_{b}^{\nu)}\delta_{c}^{(\eta}\delta_{d}^{i_{1}}\delta_{j_{2}}^{i_{2}}\cdots\delta_{j_{2h}}^{i_{2h}}\delta_{j_{2h+1}}^{i_{2h+1})}\biggr\}
\end{multline}
We have to verify that $\langle \left(B_{p}\right)_{\alpha}^{\eta}\rangle=0$ holds. Inserting Eq.~\eqref{derivordsupsimm} in the l.h.s. of Eq.~\eqref{formuladerivatesup} that, in the weak field limit up to the order $h^{2}$ becomes

\begin{multline}\label{bassen}
\sum_{h=1}^{p}\sum_{j=0}^{2h}\sum_{m=j+1}^{2h+1}\left(-1\right)^{j}\partial_{i_{0}\cdots i_{j}}\left[\sqrt{-g}a_{h}R\frac{\partial\Box^{h}R}{\partial g_{\mu\nu,\eta i_{1}\cdots i_{m}}}\right]g_{\mu\nu,i_{j+1}\cdots i_{m}\alpha}\\
\stackrel{h^{2}} =\sum_{h=1}^{p}\sum_{j=0}^{2h}\left(-1\right)^{j}\sqrt{-g}^{\left(0\right)}a_{h}\partial_{i_{0}\cdots i_{j}}R^{\left(1\right)}\eta^{j_{2}j_{3}}\cdots \eta^{j_{2h}j_{2h+1}}\eta^{ab}\eta^{cd}\biggl\{\delta_{a}^{(\mu}\delta_{d}^{\nu)}\delta_{c}^{(\eta}\delta_{b}^{i_{1}}\delta_{j_{2}}^{i_{2}}\cdots\delta_{j_{2h}}^{i_{2h}}\delta_{j_{2h+1}}^{i_{2h+1})}\\
-\delta_{a}^{(\mu}\delta_{b}^{\nu)}\delta_{c}^{(\eta}\delta_{d}^{i_{1}}\delta_{j_{2}}^{i_{2}}\cdots\delta_{j_{2h}}^{i_{2h}}\delta_{j_{2h+1}}^{i_{2h+1})}\biggr\}h_{\mu\nu,i_{j+1}\cdots i_{2h+1}\alpha}\\
=\sum_{h=1}^{p}\sum_{j=0}^{2h}\left(-1\right)^{j}a_{h}\partial_{i_{0}\cdots i_{j}}R^{\left(1\right)}Q_{\left(\mu\nu\right)}^{\ \ \ \left(\eta i_{1}\cdots i_{2h+1}\right)}h^{\mu\nu}_{\ \ ,i_{j+1}\cdots i_{2h+1}\alpha} 
\end{multline}
with
\begin{equation*}
Q_{\left(\mu\nu\right)}^{\ \ \ \left(\eta i_{1}\cdots i_{2h+1}\right)}=\frac{1}{2!\left(2h+2\right)!}\sum_{ \substack{\mu\nu\in \sigma\left({\mu\nu}\right) \\ \eta i_{1}\cdots i_{2h+1}\in\sigma\left(\eta i_{1}\cdots i_{2h+1}\right)}}Q_{\mu\nu}^{\ \ \ \eta i_{1}\cdots i_{2h+1}}
\end{equation*}
and 
\begin{equation*}
Q_{\left(\mu\nu\right)}^{\ \ \ \left(\eta i_{1}\cdots i_{2h+1}\right)}=\delta_{(\mu}^{(\eta}\delta_{\nu)}^{i_{1}}\eta^{i_{2}i_{3}}\cdots\eta^{i_{2h}i_{2h+1})}-\eta_{(\mu\nu)}\eta^{(\eta i_{1}}\eta^{i_{2}i_{3}}\cdots\eta^{i_{2h}i_{2h+1})}
\end{equation*}
where $\sigma{\left(\mu\nu\right)}$ and $\sigma{\left(\eta i_{1}\cdots i_{2h+1}\right)}$ represent the set of index permutations in the brackets. Averaging Eq.~\eqref{bassen} by fixing $\mathbf{k}$ over a suitable spacetime region adopting a harmonic gauge, we get 
\begin{multline}\label{media2}
\langle\sum_{h=1}^{p}\sum_{j=0}^{2h}\left(-1\right)^{j}a_{h}\partial_{i_{0}\cdots i_{j}}R^{\left(1\right)}Q_{\left(\mu\nu\right)}^{\ \ \ \left(\eta i_{1}\cdots i_{2h+1}\right)}h^{\mu\nu}_{\ \ ,i_{j+1}\cdots i_{2h+1}\alpha}\rangle\\
=\sum_{h=1}^{p}\sum_{j=0}^{2h}\frac{1}{2!\left(2h+2\right)!}\left(-1\right)^{j}a_{h}\sum_{ \substack{\mu\nu\in \sigma\left({\mu\nu}\right) \\ \eta i_{1}\cdots i_{2h+1}\in\sigma\left(\eta i_{1}\cdots i_{2h+1}\right)}}\langle\partial_{i_{0}\cdots i_{j}}R^{\left(1\right)}Q_{\mu\nu}^{\ \ \ \eta i_{1}\cdots i_{2h+1}}h^{\mu\nu}_{\ \ ,i_{j+1}\cdots i_{2h+1}\alpha}\rangle
\end{multline}
The average of Eq.~\eqref{media2} is independent of index permutations in the lower and upper cases of $Q_{\mu\nu}^{\ \ \ \eta i_{1}\cdots i_{2h+1}}$, that is 
\begin{equation}\label{media3}
\langle\partial_{i_{0}\cdots i_{j}}\left(-\frac{1}{2}\Box h\right)Q_{\mu\nu}^{\ \ \ \eta i_{1}\cdots i_{2h+1}}h^{\mu\nu}_{\ \ ,i_{j+1}\cdots i_{2h+1}\alpha}\rangle=\frac{1}{2}\sum_{m=2}^{p+2}\left(-1\right)^{j+h}\left(k_{m}^{2}\right)^{h+1}\left(k_{m}\right)^{\eta}\left(k_{m}\right)_{\alpha}\vert A_{m}\vert^{2}
\end{equation}
By substituting Eq.~\eqref{media3} in Eq.~\eqref{media2}, we get 
\begin{multline}\label{media4}
\langle\sum_{h=1}^{p}\sum_{j=0}^{2h}\sum_{m=j+1}^{2h+1}\left(-1\right)^{j}\partial_{i_{0}\cdots i_{j}}\left[\sqrt{-g}a_{h}R\frac{\partial\Box^{h}R}{\partial g_{\mu\nu,\eta i_{1}\cdots i_{m}}}\right]g_{\mu\nu,i_{j+1}\cdots i_{m}\alpha}\rangle\\
\stackrel{h^{2}}=\sum_{h=1}^{p}\sum_{j=0}^{2h}\left(-1\right)^{j}a_{h}\sum_{m=2}^{p+2}\left(-1\right)^{j+h}\left(k_{m}^{2}\right)^{h+1}\left(k_{m}\right)^{\eta}\left(k_{m}\right)_{\alpha}\vert A_{m}\vert^{2}\\
=\sum_{h=1}^{p}\sum_{m=2}^{p+2}\left(h+\frac{1}{2}\right)a_{h}\left(-1\right)^{h}\left(k_{m}^{2}\right)^{h+1}\left(k_{m}\right)^{\eta}\left(k_{m}\right)_{\alpha}\vert A_{m}\vert^{2}
\end{multline}
Averaging the right term in Eq.~\eqref{formuladerivatesup}, we have
\begin{multline}\label{media5}
\langle\frac{1}{4}\sum_{h=1}^{p}a_{h}\Box h \Box^{h} h^{,\eta}_{\ \ \alpha}+\frac{1}{2}\sum_{h=0}^{1}\sum_{j=h}^{p-1+h}\sum_{m=j+1-h}^{p}\left(-1\right)^{h}a_{m}\Box^{m-j}\left(h^{\eta\lambda}-\eta^{\eta\lambda}h\right)_{,i_{h}\alpha}\Box^{j+1-h}h_{,\lambda}^{\ \ i_{h}}\rangle\\
=\sum_{h=1}^{p}\sum_{m=2}^{p+2}\left(h+\frac{1}{2}\right)a_{h}\left(-1\right)^{h}\left(k_{m}^{2}\right)^{h+1}\left(k_{m}\right)^{\eta}\left(k_{m}\right)_{\alpha}\vert A_{m}\vert^{2}
\end{multline}
Finally, by averaging in the weak field limit Eq. ~\eqref{formuladerivatesup} and from Eqs.~\eqref{media4} and \eqref{media5}, we obtain:
\begin{equation}
\langle \left(B_{p}\right)_{\alpha}^{\eta}\rangle=0
\end{equation}
A similar argument gives $\langle \left(A_{p}\right)_{\alpha}^{\eta}\rangle=0$. It is 
\begin{multline}
\langle\sum_{h=1}^{p}\sum_{q=0}^{2h+1}\left(-1\right)^{q}\partial_{i_{0}\cdots i_{q}}\left[\sqrt{-g}a_{h}R\frac{\partial\Box^{h}R}{\partial g_{\mu\nu,\eta i_{1}\cdots i_{q}}}\right]g_{\mu\nu,\alpha}\rangle\\
\stackrel{h^{2}}=\frac{1}{2}\sum_{h=1}^{p}\sum_{m=2}^{p+2}a_{h}\left(-1\right)^{h+1}\left(k_{m}^{2}\right)^{h+1}\left(k_{m}\right)^{\eta}\left(k_{m}\right)_{\alpha}\vert A_{m}\vert^{2}
\end{multline}
\begin{multline}
\langle\frac{1}{2}\sum_{h=1}^{p}a_{h}\Box^{h+1}h_{,\lambda}\left(h^{\eta\lambda}-\eta^{\eta\lambda}h\right)_{,\alpha}\rangle\stackrel{h^{2}}=\frac{1}{2}\sum_{h=1}^{p}\sum_{m=2}^{p+2}a_{h}\left(-1\right)^{h+1}\left(k_{m}^{2}\right)^{h+1}\left(k_{m}\right)^{\eta}\left(k_{m}\right)_{\alpha}\vert A_{m}\vert^{2}
\end{multline}
and then averaging Eq. ~\eqref{formuladermax} on the l.h.s. and r.h.s., in the weak field limit, we have
\begin{equation}
\langle \left(A_{p}\right)_{\alpha}^{\eta}\rangle=0
\end{equation}
that completes our demonstration.
\subsection[\appendixname \thesubsection]{The polarizations of gravitational waves}\label{B}
The six polarizations in the solution \eqref{GW1} can be defined in a suitable matrix base. That is \cite{CCC}
\begin{align*}
 \epsilon_{\mu\nu}^{\left(+\right)}&=\frac{1}{\sqrt{2}}
\begin{pmatrix}
0 & 0 & 0 & 0\\0 & 1 & 0 & 0\\0 & 0 & -1 & 0\\0 & 0 & 0 & 0
\end{pmatrix}&
 \epsilon_{\mu\nu}^{\left(\times\right)}&=\frac{1}{\sqrt{2}}
\begin{pmatrix}
0 & 0 & 0 & 0\\0 & 0 & 1 & 0\\0 & 1 & 0 & 0\\0 & 0 & 0 & 0
\end{pmatrix}\\
\epsilon_{\mu\nu}^{\left(\text{TT}\right)}&=\qquad
\begin{pmatrix}
1 & 0 & 0 & 0\\0 & 0 & 0 & 0\\0 & 0 & 0 & 0\\0 & 0 & 0 & 0
\end{pmatrix}&
\epsilon_{\mu\nu}^{\left(\text{TS}\right)}&=\frac{1}{\sqrt{2}}
\begin{pmatrix}
0 & 0 & 0 & 1\\0 & 0 & 0 & 0\\0 & 0 & 0 & 0\\1 & 0 & 0 & 0
\end{pmatrix}\\
\epsilon_{\mu\nu}^{\left(1\right)}&=\frac{1}{\sqrt{2}}
\begin{pmatrix}
0 & 0 & 0 & 0\\0 & 1 & 0 & 0\\0 & 0 & 1 & 0\\0 & 0 & 0 & 0
\end{pmatrix}&
\epsilon_{\mu\nu}^{\left(L\right)}&=\qquad
\begin{pmatrix}
0 & 0 & 0 & 0\\0 & 0 & 0 & 0\\0 & 0 & 0 & 0\\0 & 0 & 0 & 1
\end{pmatrix}
\end{align*}
The $+$ and $\times$ are the two standard of general relativity. The other are related to the position of non-null terms with respect to the trace (T).
See also \cite{arturo1} for another derivation in fourth order gravity.

\end{document}